\documentclass[preprint,showpacs,preprintnumbers,amsmath,amssymb]{revtex4}
\usepackage{amsmath}

\usepackage{graphicx}
\usepackage{dcolumn}
\usepackage{bm}
\usepackage{amsmath} 
\usepackage{subfig}

\newcommand{\qed}{\nobreak \ifvmode \relax \else
      \ifdim\lastskip<1.5em \hskip-\lastskip
      \hskip1.5em plus0em minus0.5em \fi \nobreak
      \vrule height0.75em width0.5em depth0.25em\fi}

\begin{document}

\preprint{}

\title{Two-Qubit Separability Probabilities as Joint Functions of the Bloch Radii of the Qubit Subsystems}
\author{Paul B. Slater}
 \email{slater@kitp.ucsb.edu}
\affiliation{%
University of California, Santa Barbara, CA 93106-4030\\
}
\date{\today}
            
\begin{abstract}
We detect a certain pattern of behavior of separability probabilities 
$p(r_A,r_B)$ for two-qubit systems endowed with Hilbert-Schmidt, and more generally, random induced measures, where $r_A$ and $r_B$ are the Bloch radii 
($0 \leq r_A,r_B \leq 1$) of the qubit reduced states ($A,B$). We observe a relative repulsion of radii effect, that is $p(r_A,r_A) < p(r_A,1-r_A)$, except for  rather  narrow ``crossover'' intervals $[\tilde{r}_A,\frac{1}{2}]$. 
Among the seven specific cases we study are, firstly, the ``toy'' seven-dimensional $X$-states model and, then, the fifteen-dimensional two-qubit states obtained by tracing over the pure states in 
$4 \times K$-dimensions, for 
$K=3, 4, 5$, with  $K=4$ corresponding to  Hilbert-Schmidt (flat/Euclidean) measure. We also examine the real (two-rebit) $K=4$, the $X$-states $K=5$, and Bures (minimal monotone)--for which no nontrivial crossover behavior is observed--instances. In the two 
$X$-states cases, we derive analytical results; for $K=3, 4$, we propose formulas that well-fit our numerical results; and for the other  scenarios, rely presently upon
large numerical analyses. The separability probability 
crossover regions found expand in length  (lower $\tilde{r}_A$) as $K$ increases.
This report continues our efforts (arXiv:1506.08739) to extend the recent work of Milz and Strunz ({\it J. Phys. A}: {\bf{48}}  [2015] 035306) from a univariate ($r_A$) 
framework---in which they found separability probabilities to hold constant 
with $r_A$---to a bivariate ($r_A,r_B$) one. We also  analyze
the two-{\it qutrit} and {\it qubit}-qutrit counterparts reported in arXiv:1512.07210 
in this context, and study two-qubit separability probabilities of the form $p(r_A,\frac{1}{2})$. A physics.stack.exchange link to a contribution by Mark Fischler 
addressing, in considerable detail, the construction of suitable bivariate distributions is indicated at the end of the paper.
\end{abstract}

\keywords{$2 \cdot 2$ quantum systems,  Bloch sphere, Bloch ball,   Milz-Strunz, reduced density matrix, two qubits,  Hilbert-Schmidt measure,  separability probabilities, $X-states, random induced measure, Bures measure, two-rebits, two-qubits, bivariate probability distributions, two-qutrits, qubit-qutrit, qubit-qudit}

\maketitle

\tableofcontents
\section{Introduction}
\begin{quote}
``The Bloch sphere provides a simple representation for the state space of the most primitive 
quantum unit--the qubit--resulting in geometric intuitions that are invaluable in countless fundamental information-processing scenarios'' \cite{Jevtic}.
\end{quote}

Motivated by recent interesting work of Milz and Strunz \cite{milzstrunz}, indicating the constancy of Hilbert-Schmidt two-qubit (and qubit-qutrit) separability probabilities over the Bloch radius of qubit subsystems, we began a study in \cite{Repulsion} devoted to extending their ``single-Bloch radius'' ($r_A$) results to ``joint-Bloch-radii" ($r_A,r_B$) analyses (cf. \cite{Gamel}). Most of the many results/figures
reported in \cite{Repulsion} were based on extensive numerical investigations. However, a set of exact results was obtained for the ``toy'' model of $X$-states \cite{Xstates2}, that is  X-patterned $4 \times 4$ density matrices having zero values at the eight entries--(1,2), (1,3), (2,1), (2,4), (3,1), (3,4), (4,2) and (4,3).

Milz and Strunz had found numerically-based evidence that  the Hilbert-Schmidt (HS) volumes of the fifteen-dimensional convex sets of two-qubit systems and of their separable subsystems  were {\it both} proportional to $(1-r_A^2)^6$  \cite[eqs. (23), (30),(31)]{milzstrunz}. The consequent {\it constant} ratio (separability probability) of the two (simply proportional) volume functions appeared to be $\frac{8}{33}$--a remarkably simple value for which a large body of diverse support had already been developed \cite{slater833,MomentBased,slaterJModPhys,FeiJoynt,WholeHalf,Dubna} 
\cite[sec. VII]{Fonseca-Romero} \cite[sec. 4]{Shang}, though yet no formal proof. 
(Let us note, however, that Lovas and Andai have recently reported substantial advances in this direction. They proved the $\frac{29}{64}$ two-rebit counterpart conjecture, and presented ``an integral formula...which hopefully will help to prove the 
$\frac{8}{33}$ result'' \cite{lovasandai}.)
\section{$X$-states analyses}
\subsection{Hilbert-Schmidt ($K=4$) case} \label{XK=4}
For the  $X$-states, occupying a seven-dimensional subspace of the full fifteen-dimensional space, it was possible for Milz and Strunz to {\it formally} demonstrate that the counterpart total and separable volume functions, similarly, were both  again proportional, but now to $(1-r_A^2)^3$ (the square root of
the fifteen-dimensional result). The corresponding constant (but at the isolated pure states [$r_A=1$] boundary) HS separability probability was greater than $\frac{8}{33}$, that is 
$\frac{2}{5}$ \cite[Apps. A, B]{milzstrunz}. This  $\frac{2}{5}$ result was also subsequently 
proven  in \cite{LatestCollaboration}, along with companion $X$-states findings for the broader class of random induced measures \cite{Induced,aubrun2,adachi2009random}. (A distinct analytical approach, based on the Cholesky decomposition of density matrices, was utilized.)

\begin{figure}
\includegraphics{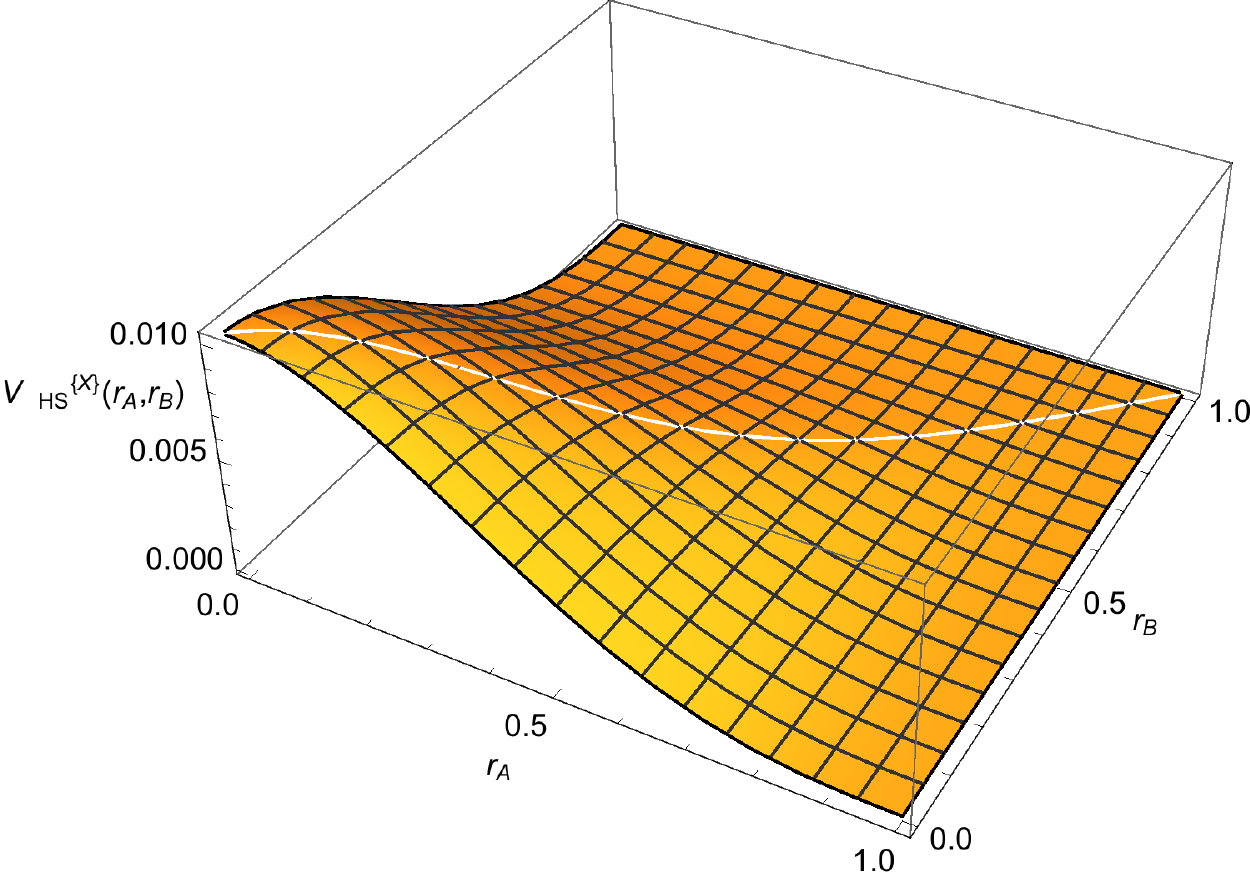}
\caption{\label{fig:Xbivariate}Bivariate Hilbert-Schmidt volume distribution (\ref{Xtotal}) for the $X$-state model}
\end{figure}

\begin{figure}
\includegraphics{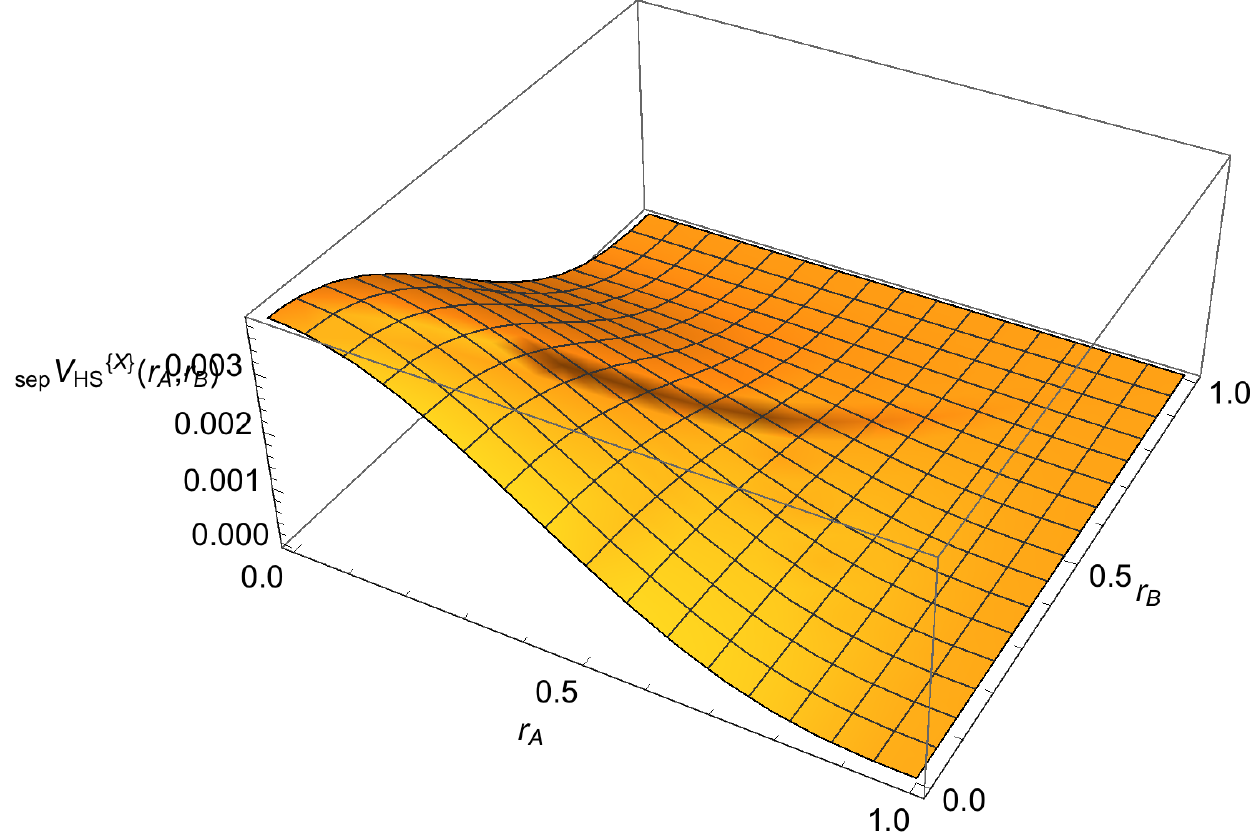}
\caption{\label{fig:XbivariateSep}Bivariate Hilbert-Schmidt 
{\it separable} volume distribution (\ref{Xsep}) for the $X$-state model}
\end{figure}

\begin{figure}
\includegraphics{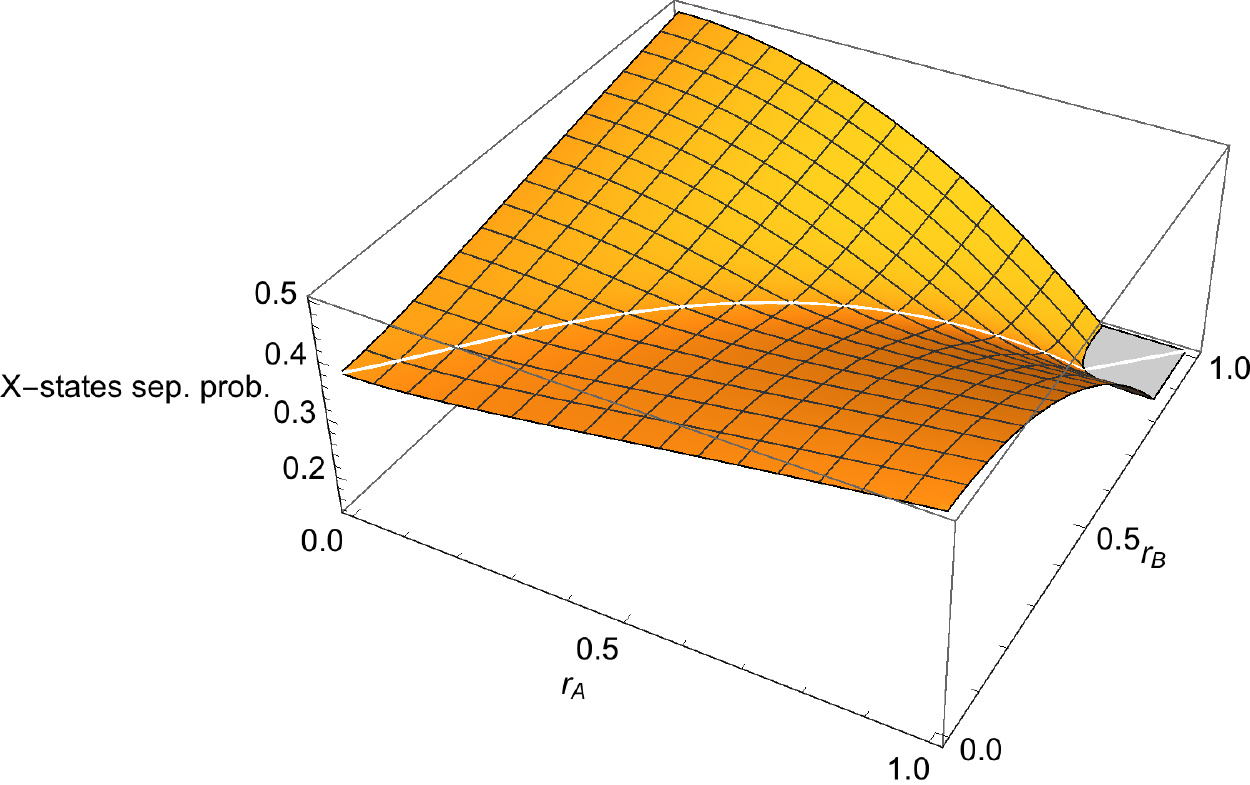}
\caption{\label{fig:XbivariateSepProb}Bivariate Hilbert-Schmidt 
separability probability distribution  (\ref{BivSepProb})--the ratio of 
Fig.~\ref{fig:XbivariateSep} to Fig.~\ref{fig:Xbivariate}--for the $X$-state model}
\end{figure}

In \cite{Repulsion}, we employed the $X$-states parametrization and transformations indicated by Braga, Souza and Mizrahi \cite[eqs. (6), (7)]{BSM}. We were able to reproduce the Hilbert-Schmidt {\it univariate} volume result of Milz and Strunz 
\cite[eq. (20), Fig. 1]{milzstrunz} \cite{LatestCollaboration},
\begin{equation}
V^{(X)}_{HS}(r) =\frac{\pi^2}{2304} (1-r^2)^3,
\end{equation}
as the marginal distribution (over either $r_A$ or $r_B$) of the {\it bivariate} distribution (Fig.~\ref{fig:Xbivariate}),
\begin{equation} \label{Xtotal} 
_{{tot}}V^{(X)}_{HS}(r_A,r_B) =
\end{equation}
\begin{displaymath}
\begin{cases}
 -\frac{1}{960} \pi ^2 \left(r_A-1\right){}^3 \left(r_A \left(r_A+3\right)-5
   r_B^2+1\right) & r_A>r_B \\
 -\frac{1}{960} \pi ^2 \left(r_B-1\right){}^3 \left(-5 r_A^2+r_B
   \left(r_B+3\right)+1\right) & r_A<r_B
\end{cases}.
\end{displaymath}
To, then, obtain the desired $X$-states bivariate separability probability distribution 
$p^{(X)}_{HS}(r_A,r_B)$, we further found the {\it separable} volume counterpart to (\ref{Xtotal}) (Fig.~\ref{fig:XbivariateSep}),
\begin{equation} \label{Xsep} 
_{{sep}}V^{(X)}_{HS}(r_A,r_B) =
\end{equation}
\begin{displaymath}
\begin{cases}
 -\frac{\pi ^2 \left(r_A-1\right){}^3 \left(5 \left(r_A+3\right) r_B^4-10 \left(3
   r_A+1\right) r_B^2+8 r_A^2+9 r_A+3\right)}{7680} & r_A>r_B \\
 -\frac{\pi ^2 \left(r_B-1\right){}^3 \left(5 r_A^4 \left(r_B+3\right)-10 r_A^2 \left(3
   r_B+1\right)+r_B \left(8 r_B+9\right)+3\right)}{7680} & r_A<r_B
\end{cases},
\end{displaymath}
and took their ratio (Fig.~\ref{fig:XbivariateSepProb}) (note the cancellation of the $(r-1)^3$-type factors),
\begin{equation} \label{BivSepProb}
p^{(X)}_{HS}(r_A,r_B)=\frac{_{{sep}}V^{(X)}_{HS}(r_A,r_B)}{_{{tot}}V^{(X)}_{HS}(r_A,r_B)}=
\end{equation}
\begin{displaymath}
\begin{cases}
 \frac{5 \left(r_A+3\right) r_B^4-10 \left(3 r_A+1\right) r_B^2+8 r_A^2+9 r_A+3}{8
   \left(r_A \left(r_A+3\right)-5 r_B^2+1\right)} & r_A>r_B \\
 \frac{5 r_A^4 \left(r_B+3\right)-10 r_A^2 \left(3 r_B+1\right)+r_B \left(8
   r_B+9\right)+3}{8 \left(-5 r_A^2+r_B \left(r_B+3\right)+1\right)} & r_A<r_B
\end{cases}.
\end{displaymath}
(Numerical integration of this function over $[0,1]^2$ yielded $0.381678 \approx 0.4$--so, it would seem that $p^{(X)}_{HS}(r_A,r_B)$ is not strictly a scaled version of 
a {\it doubly-stochastic} measure \cite{SungurNg,Ruschendorf}, as we had speculated it might be.) 

\begin{figure}
\includegraphics{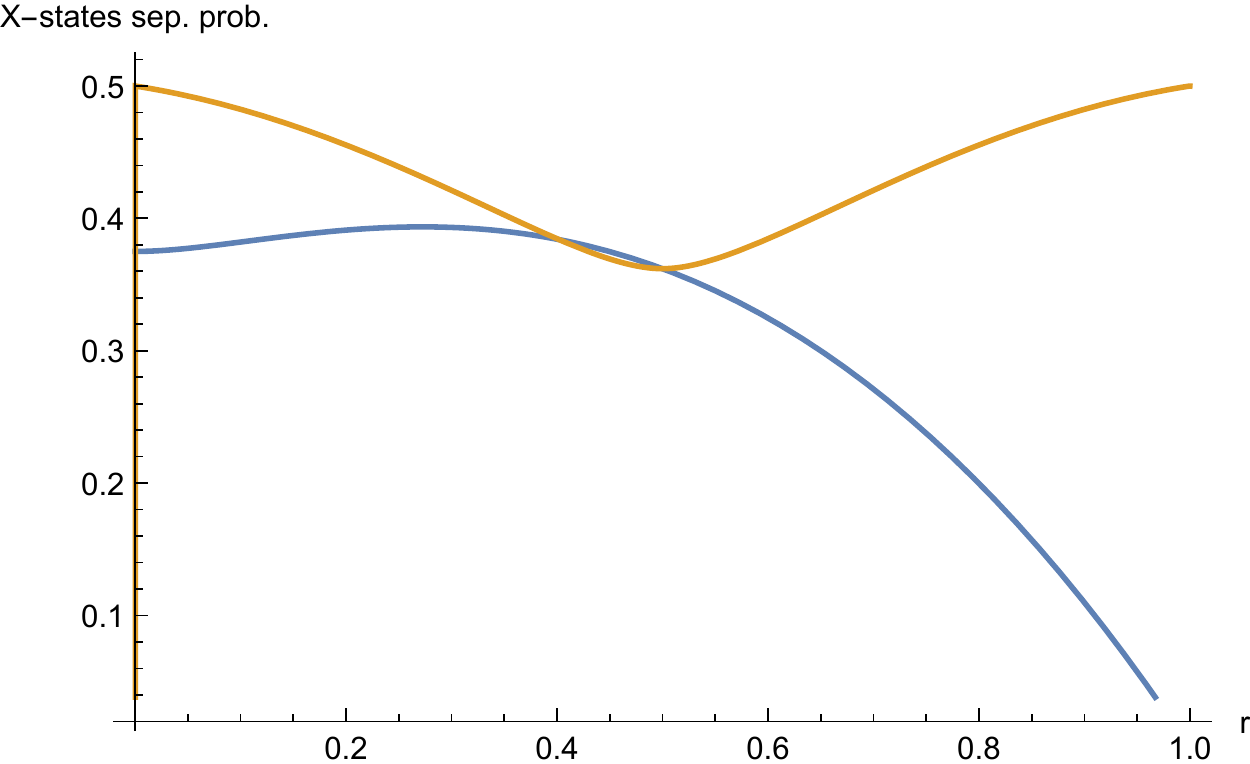}
\caption{\label{fig:XbivariateUpperLower}(Largely lower) $r_A=r_B$--given by (\ref{Lower})--and (largely upper) $r_A+r_B=1$ curves--given by (\ref{Upper})--for bivariate Hilbert-Schmidt 
$X$-states separability probability distribution. The minimum of the upper ``antidiagonal" curve is at $r_A=\frac{1}{2}$, while the maximum of the lower ``diagonal" curve is at 0.27227007. In the crossover interval $r_A \in [0.40182804, \frac{1}{2}]$, the $p^{(X)}_{HS}(r_A,1-r_A)$ curve is dominated by the $p^{(X)}_{HS}(r_A,r_A)$ curve.}
\end{figure}

\begin{figure}
\includegraphics{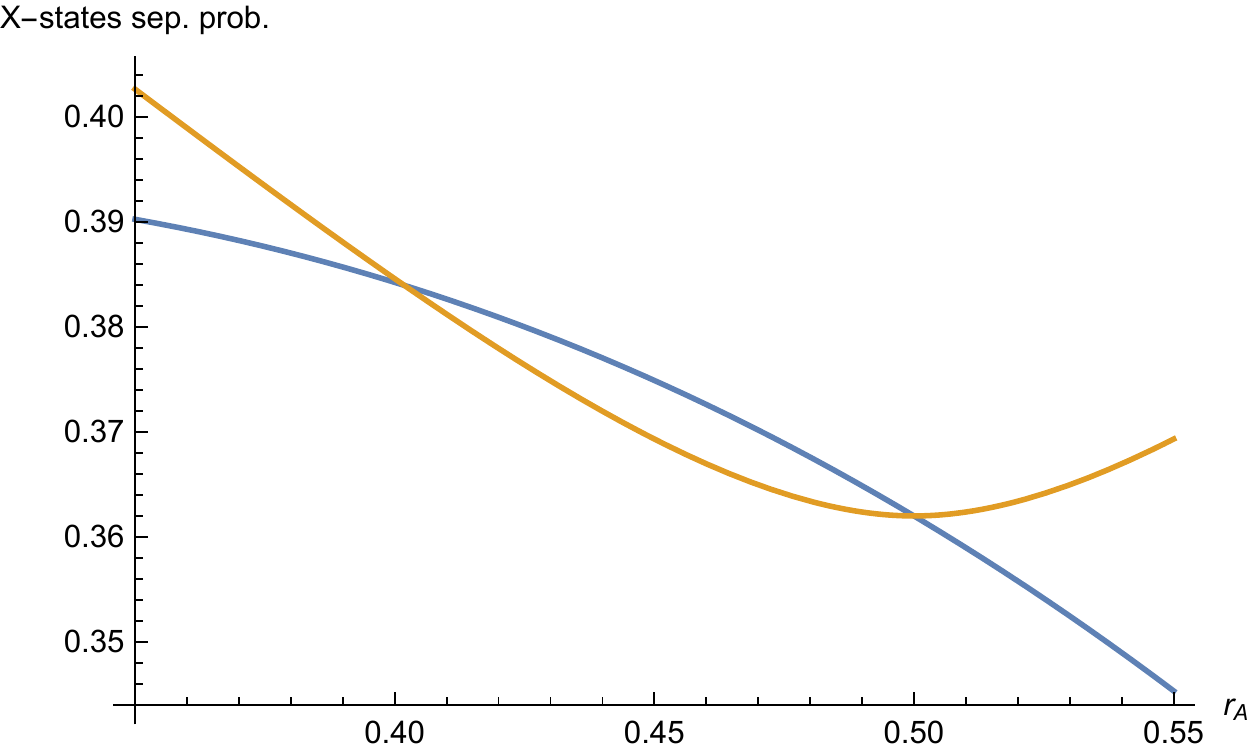}
\caption{\label{fig:RefinedXUpperLower}Closer examination of the $X$-states crossover region in 
Fig.~\ref{fig:XbivariateUpperLower}, the lower bound being $\tilde{r}_A = 0.40182804$, a root of the quintic equation (\ref{XK4})}
\end{figure}

Fig.~\ref{fig:XbivariateUpperLower}  (also \cite[Fig. 50]{Repulsion}) 
shows the (largely lower) $r_A=r_B$
and (largely upper) $r_A +r_B =1$ one-dimensional cross-sections of 
Fig.~\ref{fig:XbivariateSepProb}. (We computed the correlation between $r_A$ and $r_B$ to be $1-\frac{11206656}{37748736-10080 \pi ^2+\pi ^4} \approx 0.702341$ for all states and only slightly less, $1-\frac{74649600}{235929600-25200 \pi ^2+\pi ^4} \approx 0.68326$, for the separable $X$-states (cf. \cite{deVicente}).)
In Fig.~\ref{fig:RefinedXUpperLower} we show more closely the {\it crossover} region 
in which the $p^{(X)}_{HS}(r_A,1-r_A)$ curve becomes dominated by the $p^{(X)}_{HS}(r_A,r_A)$ curve.

The analytic form of the $r_A=r_B$ $X$-states separability probability curve is
\begin{equation} \label{Lower}
p^{(X)}_{HS}(r_A,r_A) =-\frac{(r_A-1) (5 r_A (r_A (r_A+5)+3)+3)}{32 r_A+8}.
\end{equation}
At particular points of interest, we have $p^{(X)}_{HS}(\frac{1}{2},\frac{1}{2})= 
\frac{139}{384}=
\frac{139}{2^7 \cdot 3}$, $p^{(X)}_{HS}(0,0) =\frac{3}{8}$, and  $p^{(X)}_{HS}(1,1)=0$. The maximum of  $p^{(X)}_{HS}(r_A,r_A)$ is achieved at the positive root ($r_A \approx 0.2722700792$) of the cubic
equation $3 r_A^3+9 r_A^2+r_A-1=0$. Its value ($\approx 0.393558399$) there  is the positive 
root of the cubic equation $54 r_A^3 + 108 r_A^2 - 28 r_A - 9=0$.
 
On the other hand, the minimum  of the $r_A+r_B=1$ (``antidiagonal") curve 
\begin{equation} \label{Upper}
p^{(X)}_{HS}(r_A,1-r_A) = 
\begin{cases}
 -\frac{(r_A-2) r_A \left(5 r_A \left(r_A^2+r_A-10\right)+28\right)+8}{8 (r_A (4 r_A-13)+4)} & 2 r_A>1 \\
 \frac{r_A (r_A (5 r_A ((r_A-4) r_A-6)+32)+25)-20}{8 (r_A (4 r_A+5)-5)} & 2 r_A<1
\end{cases}
\end{equation}
is, again, $\frac{139}{384}$, clearly attained at the point of 
symmetry, $r_A=\frac{1}{2}$.
(We employ the terms "diagonal" and "antidiagonal" to describe the two types of curves
under investigation, in reference to the entries of the $100 \times 100$ data matrices we employ for their estimation.)
Also, at the endpoints,
\begin{equation} \label{Xmax}
p^{(X)}_{HS}(0,1) = p^{(X)}_{HS}(1,0) =\frac{1}{2}
\end{equation}
are the two maxima of $ p^{(X)}_{HS}(r_A,1-r_A)$.

We note--in line with our general observations throughout the paper--that in the 
crossover region $r_A \in [0.40182804, \frac{1}{2}]$ (Figs.~\ref{fig:XbivariateUpperLower} and \ref{fig:RefinedXUpperLower}), the $p^{(X)}_{HS}(r_A,1-r_A)$ curve changes from  dominating  the $p^{(X)}_{HS}(r_A,r_A)$ curve to being subordinate to it--so that the radii are relatively ``attractive" and not relatively ``repulsive" in this domain. The lower bound of the region $\tilde{r}_A =0.40182804$ is
a root of the quintic equation 
(with remarkably simple coefficients)
\begin{equation} \label{XK4}
4 r_A^5+5 r_A^4-8 r_A^3-14 r_A^2+4 r_A+1=0.
\end{equation}
The maximum gap of 0.0056796160 between the two curves in the crossover region  is attained at $r_A=0.4564893379$.
\subsection{Random induced ($K=5$) case} \label{XK=5}
\begin{figure}
\includegraphics{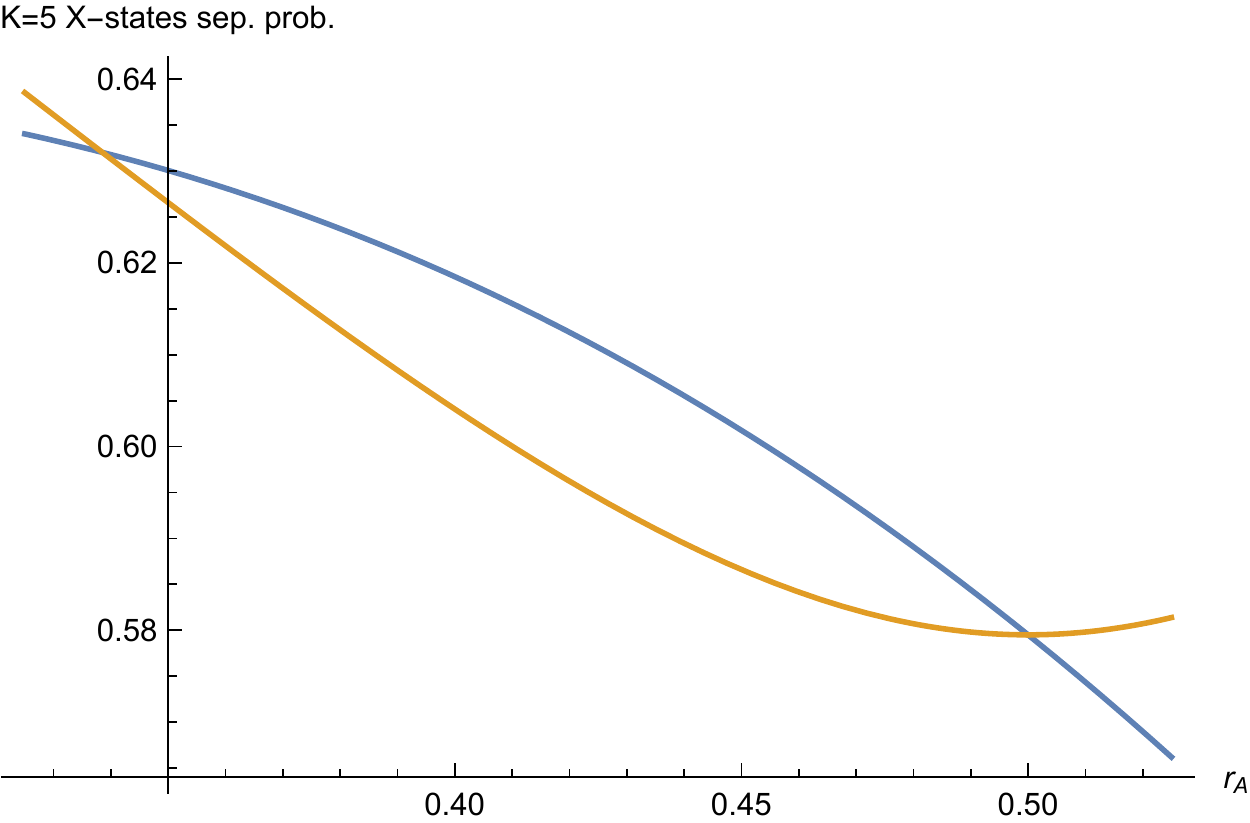}
\caption{\label{fig:RefinedXK5UpperLower}Crossover region for the random induced $K=5$ $X$-states case, the lower intersection point of the crossover region being $\tilde{r}_A = 0.3385355079$, a root of the eighth-degree equation (\ref{XK5})}
\end{figure}

Exact total and separable volume and (consequent) separability probability formulas have been  reported 
\cite[sec. IX.D]{Repulsion} also for the $X$-states random induced $K=5$ counterpart.
(The marginal total and separable volumes are now both proportional to $(1-r_A^2)^5$.)
In Fig.~\ref{fig:RefinedXK5UpperLower} we show the (more pronounced) crossover behavior in that scenario. The lower crossover point of $\tilde{r}_A = 0.3385355079$ is a root of the eighth-degree equation (with rather simple Òwell-behavedÓ coefficients--all divisible by 7, but for 27)
\begin{equation} \label{XK5}
112 r_A^8+252 r_A^7-203 r_A^6-938 r_A^5-441 r_A^4+728 r_A^3+27 r_A^2-42 r_A-7=0.
\end{equation}
Consistently with our general observations below, this lower boundary $\tilde{r}_A = 0.3385355079$ of the crossover
region is smaller than that reported above (eq. (\ref{XK4})), $\tilde{r}_A = 0.40182804$, 
in the  Hilbert-Schmidt ($K=4<5$) $X$-states scenario.
\section{Full two-qubit and two-rebit analyses}
Now, let us transition from studying these two seven-dimensional $X$-states examples, to five--$K=3, 4, 5$, rebit, and Bures cases--for the full fifteen-dimensional two-qubit states. In all these cases we generated corresponding sets of random density matrices, and
discretized the values of the two Bloch radii found into intervals of length 
$\frac{1}{100}$,
obtaining thereby $100 \times 100$ data matrices of separable and total counts.
\subsection{Random induced ($K=3)$ case} \label{FullK3}
Firstly, we study the instance when this set is endowed with the $K=3$ instance of random induced measure \cite{Induced,ingemarkarol,adachi2009random}. (The corresponding [overall] separability probability, then, appears to be $\frac{1}{14} \approx 0.0714286$ 
\cite[eq. (2)]{LatestCollaboration} \cite[Fig. 17]{Repulsion}.). The (apparent, well-fitting) total volume formula we obtained, after extensive investigations, was
\begin{equation} \label{totalK3}
_{{tot}}V_{K=3}(r_A,r_B) =
\begin{cases}
 \frac{8 \left(r_A-1\right){}^4 \left(r_A^2+4 r_A-5 r_B^2\right)}{r_A} & 0<r_B<r_A<1 \\
 \frac{8 \left(r_B-1\right){}^4 \left(r_B \left(r_B+4\right)-5 r_A^2\right)}{r_B} &
   0<r_A<r_B<1 \\
 -32 \left(r_A-1\right){}^5 & r_B=r_A
\end{cases},
\end{equation}
choosing to normalize so that $_{{tot}}V_{(K=3)}(\frac{1}{2},\frac{1}{2})=1$.

Further, for $r_B=r_A$, we appear to have
\begin{equation} \label{separableK3}
_{{sep}}V_{K=3}(r_A,r_A) =
\left(r_A-1\right){}^6 \left(r_A^2+6 r_A+1\right),
\end{equation}
so that, by taking a ratio, we obtain the diagonal curve
\begin{equation} \label{K3diagonal}
p_{K=3}(r_A,r_A)= \frac{1}{32} \left(1-r_A\right) \left(r_A^2+6 r_A+1\right).
\end{equation}
For $0<r_B<r_A<1$,
\begin{equation} \label{separableK3B}
_{{sep}}V_{K=3}(r_A,r_B) =
\end{equation}
\begin{displaymath}
-\frac{\left(r_A-1\right){}^4 \left(-r_A^3 \left(6012 r_B+2351\right)+2 r_A^2 \left(2424
   r_B-9859\right)+T r_A+2785 r_A^4+S r_B\right)}{5100 r_A}
\end{displaymath}
with
\begin{displaymath}
S=-22675 r_B^3-852 r_B^2+470 r_B+96; \hspace{.3in} T = -5100 r_B^4+5502 r_B^3+49355 r_B^2-1152 r_B-5196.
\end{displaymath}
The $0<r_A<r_B<1$ component of this piecewise function 
$_{{sep}}V_{K=3}(r_A,r_B)$ can be obtained
by interchanging the roles of $r_A$ and $r_B$ in (\ref{separableK3B}). 
(These formulas [for which we lack formal proofs] were developedÑÑ-with very considerable, diverse fitting effortsÑÑ-using 10,962,000,000 randomly generated $4 \times 4$ density matrices assigned $K=3$ measure, employing the 
Ginibre-matrix-based algorithm specified in \cite{Miszczak} (cf. \cite{Miszczak2})). 

The marginal distribution of the total volume function (\ref{totalK3}) over $r_B$ (cf. \cite[eq. (24)]{milzstrunz}) is 
$\frac{4 \pi}{3} \left(r_A^2-1\right){}^4$, and of the separable volume function, 
$\frac{2 \pi}{21} \left(r_A^2-1\right){}^4$, giving usÑÑ-taking their ratioÑÑ-the {\it constant} separability probability for this scenario of $\frac{1}{14}=
(\frac{2 \pi}{21})/(\frac{4 \pi}{3})$ 
\cite[eq. (2)]{LatestCollaboration} \cite[Fig. 17]{Repulsion}.

Further, we have the antidiagonal function
\begin{equation} \label{K3antidiagonal}
p_{K=3}(r_A,1-r_A)=
\begin{cases}
 \frac{5100 r_A^5-24480 r_A^4-66682 r_A^3+49256 r_A^2+38325 r_A-24480}{40800 \left(4
   r_A^2+6 r_A-5\right)} & 0<r_A<\frac{1}{2} \\
 \frac{-5100 r_A^5+1020 r_A^4+113602 r_A^3-246670 r_A^2+135629 r_A-22961}{40800 \left(4
   r_A^2-14 r_A+5\right)} & \frac{1}{2}<r_A<1
\end{cases}.
\end{equation}
Its maximum is attained at the two endpoints of [0,1] (cf. (\ref{Xmax}))
\begin{equation} 
p_{K=3}(0,1) = p_{K=3}(1,0) =\frac{1}{2}.
\end{equation}

\begin{figure}
\includegraphics{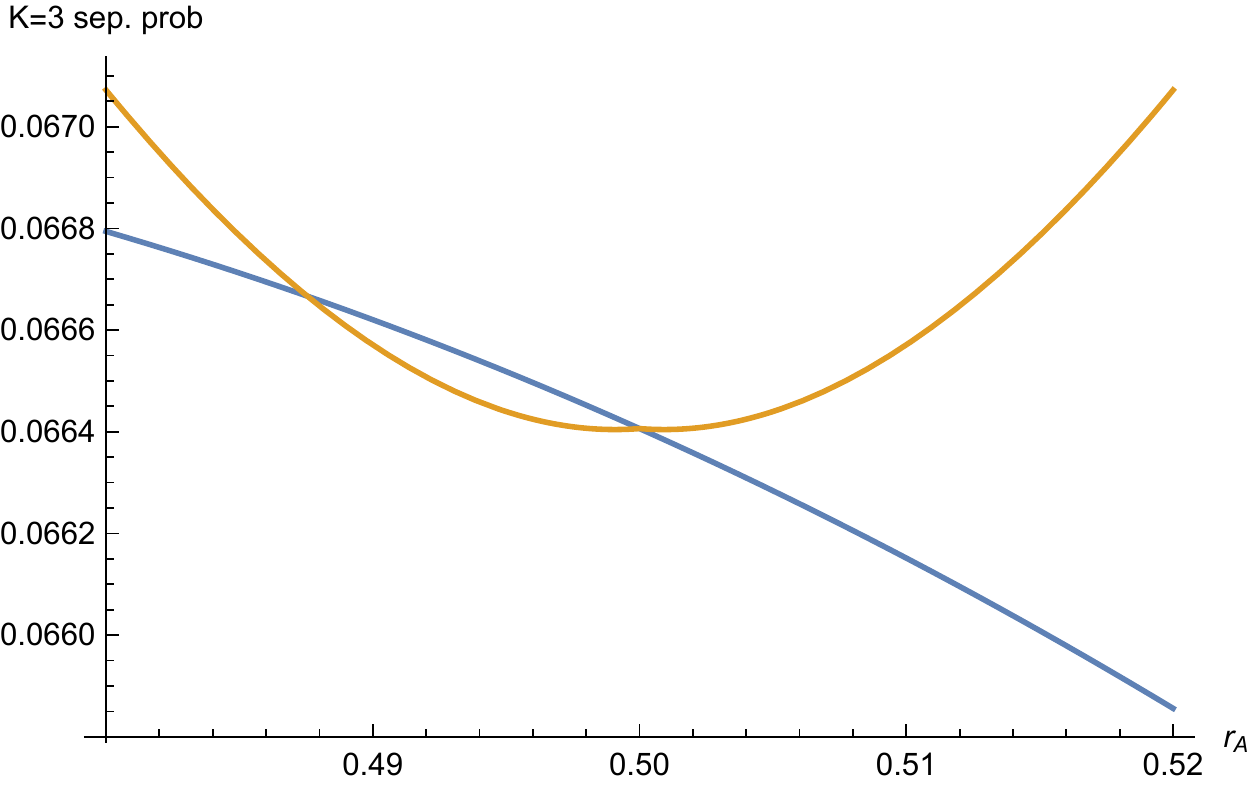}
\caption{\label{fig:RefinedK3UpperLower}Crossover region for $K=3$, based upon the separability probability formulas ((\ref{K3diagonal}), (\ref{K3antidiagonal})), with 
the lower intersection point of the crossover region being $\tilde{r}_A= 0.48754$}
\end{figure}

Based upon these $K=3$ volume formulas ((\ref{totalK3}), (\ref{separableK3})), we find, solving the quartic equation, 
\begin{equation}
5100 r_A^4+6885 r_A^3-26711 r_A^2-26340 r_A+18105=0,
\end{equation}
that the lower boundary is $\tilde{r}_A =0.487543066126$, rather near to 
$r_A =\frac{1}{2}$.
(Alternatively, interpolating the raw data, 
we obtain an estimate $\tilde{r}_A =0.488124$.)
In Fig.~\ref{fig:RefinedK3UpperLower} (cf. Fig.~\ref{fig:RefinedXUpperLower}) we show this crossover region. 
\subsection{Hilbert-Schmidt ($K=4$) case}
\begin{figure}
\includegraphics{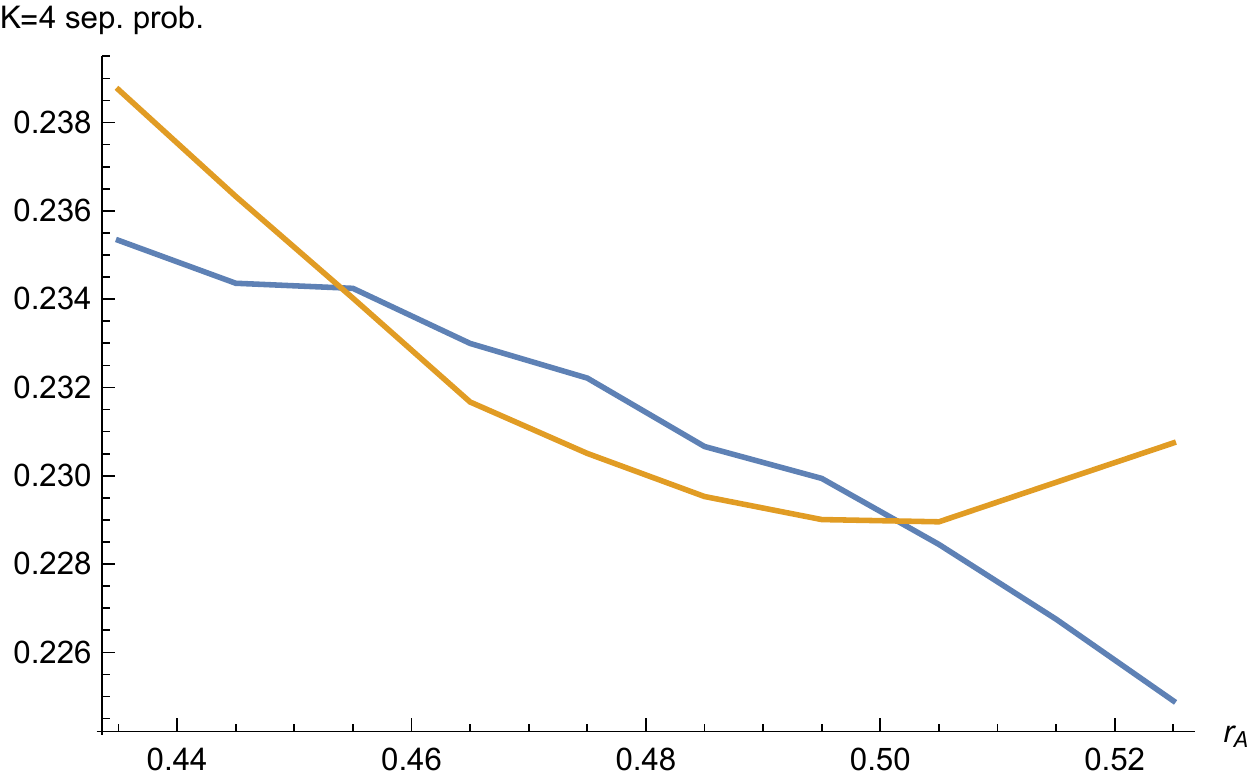}
\caption{\label{fig:RefinedK4UpperLower}Crossover region for the Hilbert-Schmidt case $K=4$, based upon 13,800,000,000 randomly generated density matrices, with the sampled Bloch radii $r_A,r_B \in [0,1]$ discretized into intervals of length $\frac{1}{100}$. Our estimate of the lower crossover point is $\tilde{r}_A=0.453893$. The maximum gap of 0.001708 between the two curves in the crossover region  is attained at $r_A=0.474381$.}
\end{figure}

In Fig.~\ref{fig:RefinedK4UpperLower}, we show the crossover behavior for the 
fundamental Hilbert-Schmidt $K=4$ case. 
(To reiterate, a considerable body of strongly compelling evidence has been developed that the corresponding separability probability is $\frac{8}{33} \approx 0.242424$
\cite{slater833,MomentBased,slaterJModPhys,FeiJoynt,WholeHalf,Dubna} 
\cite[sec. VII]{Fonseca-Romero} \cite[sec. 4]{Shang}.) 

\begin{figure}
\includegraphics{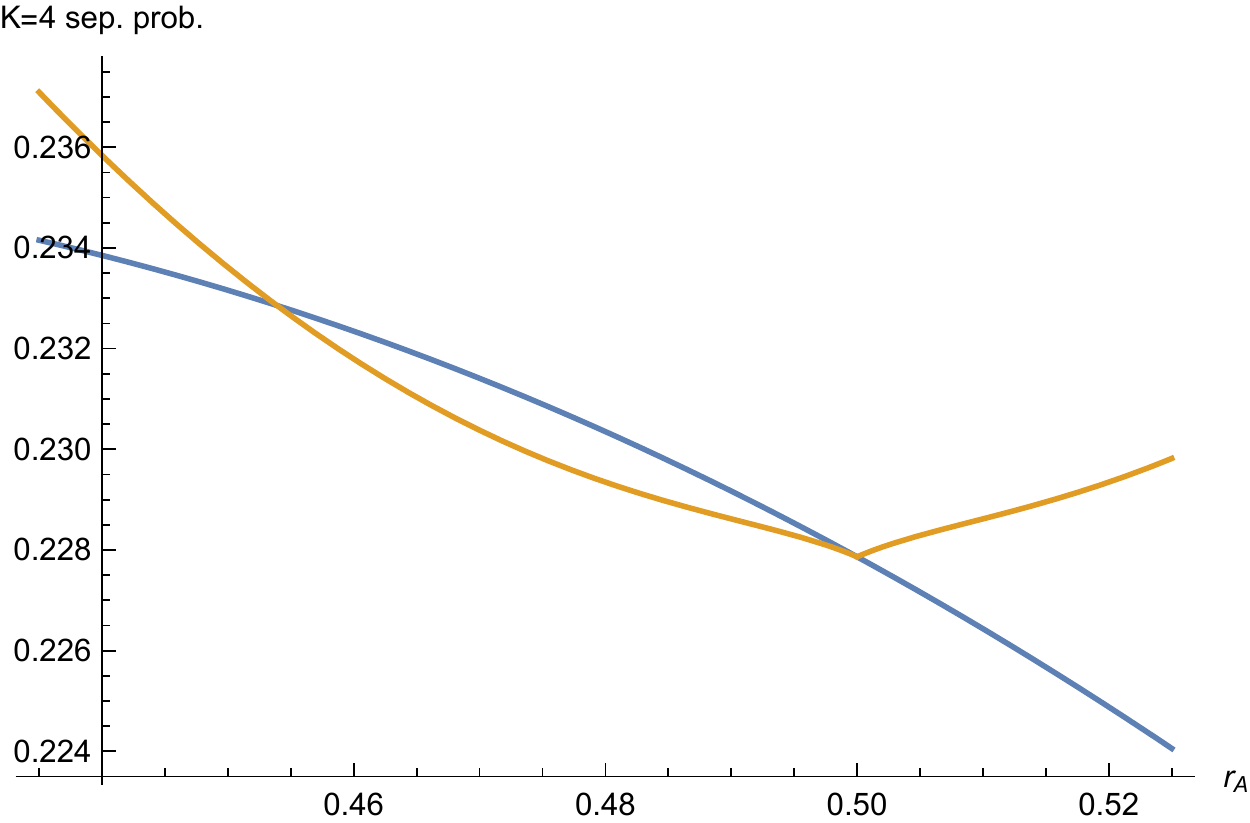}
\caption{\label{fig:TheoreticalK4}Predicted crossover region (cf. Fig.~\ref{fig:RefinedK4UpperLower}) for the Hilbert-Schmidt case $K=4$,
based upon the formulas ((\ref{K4diagonal}), (\ref{K4antidiagonal})) 
for $p^{K=4}_{HS}(r_A,r_A)$ and $p^{K=4}_{HS}(r_A,1-r_A)$}
\end{figure}

Choosing again to normalize so that $_{{tot}}V^{K=4}_{HS}(\frac{1}{2},\frac{1}{2})=1$, it appears that
\begin{equation}
_{{tot}}V^{K=4}_{HS}(r_A,r_A) =\frac{256}{5} \left(1-r_A\right){}^8 \left(8 r_A+1\right)
\end{equation}
and
\begin{equation}
_{{sep}}V^{K=4}_{HS}(r_A,r_A) =\frac{28}{3} \left(1-r_A\right){}^9 \left(29 r_A^2+\frac{17 r_A}{2}+1\right),
\end{equation}
so that, a fit to the diagonal curve can be obtained using
\begin{equation} \label{K4diagonal}
p^{K=4}_{HS}(r_A,r_A) = -\frac{35 \left(r_A-1\right) \left(58 r_A^2+17 r_A+2\right)}{384 \left(8 r_A+1\right)}.
\end{equation}
Further, for the antidiagonal curve, we have a close fit (using a chi-squared objective function) for the region $r_A \in [0,\frac{1}{2}]$ (the curve for $r_A \in [\frac{1}{2},1]$ can be obtained by 
replacing $r_A$ by $1-r_A$),
\begin{equation} \label{K4antidiagonal}
p^{K=4}_{HS}(r_A,1-r_A) \approx 
\end{equation}
\begin{displaymath}
\frac{-0.660807 r_A^6-119.919 r_A^5+237.198 r_A^4-200.68 r_A^3+90.0466 r_A^2-21.6016
   r_A+2.32483}{-1. r_A^4-66.164 r_A^3+75.933 r_A^2-30.4436 r_A+4.64965}.
\end{displaymath}
In Fig.~\ref{fig:TheoreticalK4}, we show the predicted $K=4$ crossover region based on these last two formulas.
(The marginal distributions of the total and separable $K=4$ volume functions  over 
$r_B$ appear, as Milz and Strunz argued, to be {\it both} proportional to  
$ \left(r_A^2-1\right){}^6$ \cite[eq. (23)]{milzstrunz}, with the associated 
constant ratio being $\frac{8}{33}$.)
\subsection{Random induced ($K=5$) case}
\begin{figure}
\includegraphics{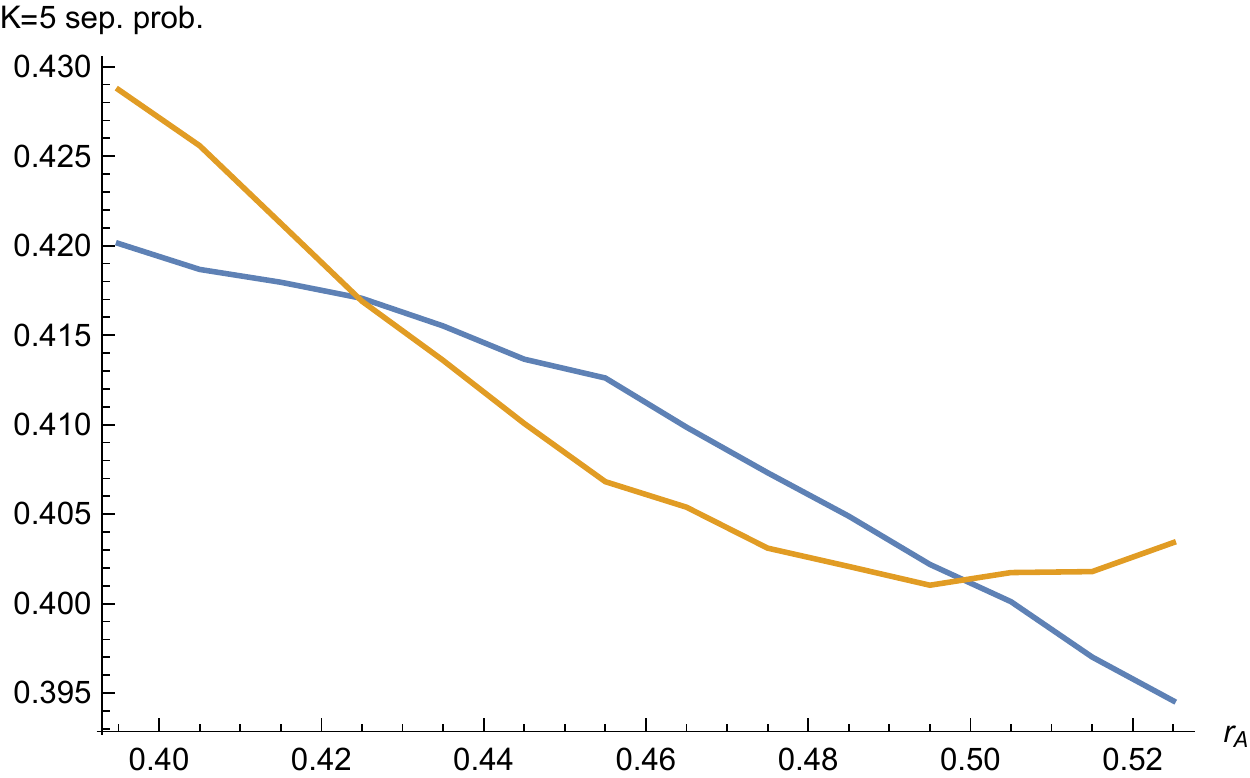}
\caption{\label{fig:RefinedK5UpperLower}Crossover region for $K=5$, based upon 6,343,000,000 randomly generated density matrices, with the sampled Bloch radii $r_A,r_B \in [0,1]$ discretized into intervals of length $\frac{1}{100}$. Our estimate of the lower crossover point is $\tilde{r}_A=0.424453$.}
\end{figure}

In Fig.~\ref{fig:RefinedK5UpperLower} we show the results for the two-qubit  
random induced $K=5$ analysis. Normalizing again so that $_{{tot}}V_{K=5}(\frac{1}{2},\frac{1}{2})=1$, it appears that \cite[sec. IV.B]{Repulsion}
\begin{equation}
_{{tot}}V_{K=5}(r_A,r_A) =\frac{4096}{33} \left(1-r_A\right){}^{11} \left(40 r_A^2+11 r_A+1\right).
\end{equation}
A good fit can be obtained using 
\begin{equation}
_{{sep}}V_{K=5}(r_A,r_A) =49 \left(1-r_A\right){}^{12} \left(108 r_A^3+\frac{111 r_A^2}{2}+10 r_A+1\right),
\end{equation}
so that
\begin{equation}
p_{K=5}(r_A,r_A) \approx -\frac{1617 \left(r_A-1\right) \left(216 r_A^3+111 r_A^2+20 r_A+2\right)}{8192 \left(40
   r_A^2+11 r_A+1\right)}.
\end{equation}
The corresponding (overall) separability probability  appears to be $\frac{61}{143} \approx 0.426573$ \cite[eq. (2), Table II]{LatestCollaboration} \cite[Fig. 24]{Repulsion}, obtainable by taking the ratio of marginal separable and total volume functions, {\it both} proportional
to $ \left(r_A^2-1\right){}^8$. So, for $K=3,4,5$, we have the sequence 
($2 K-1$) of exponents of
$ \left(r_A^2-1\right){} $ of 4, 6 and 8 for the marginal distributions.

We see that, as a general rule, the lower bounds ($\tilde{r}_A$) to the crossover regions decrease as $K$ increases.
\subsection{Two-rebit and Bures cases}
\begin{figure}
\includegraphics{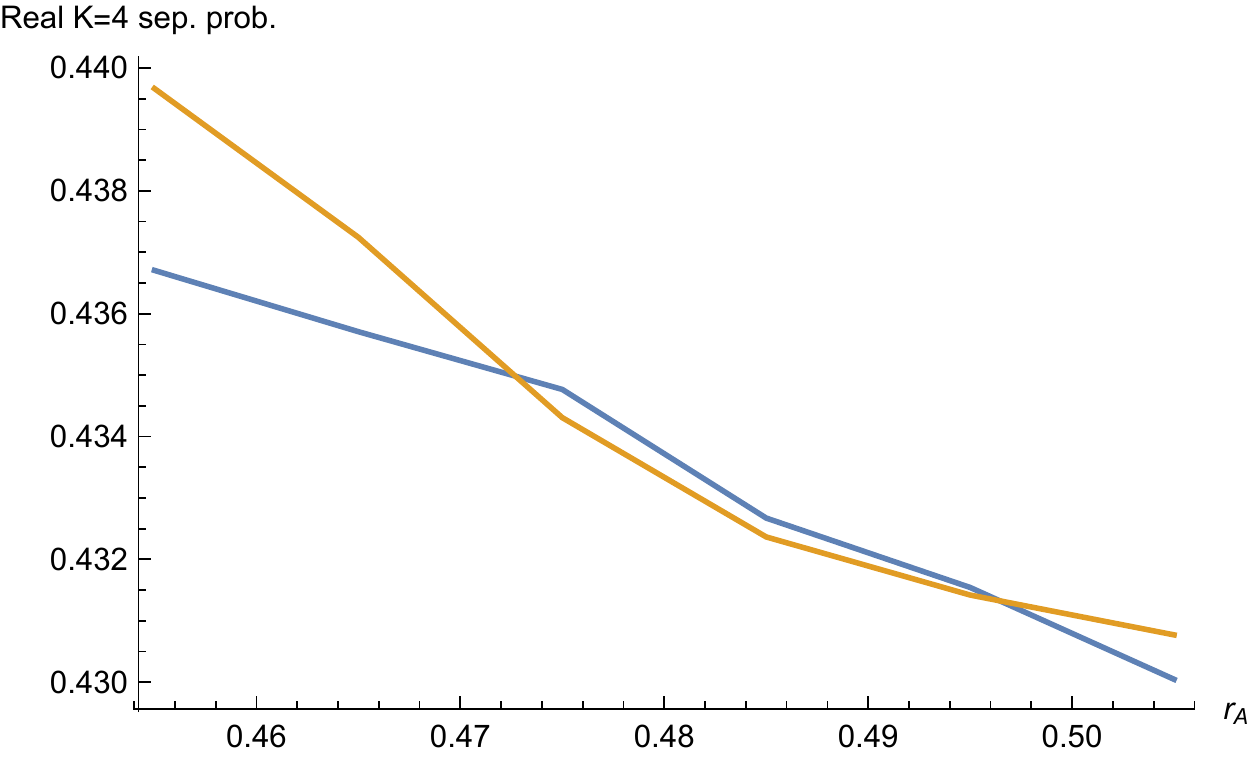}
\caption{\label{fig:RefinedRealUpperLower}Crossover region for the Hilbert-Schmidt case $K=4$, based upon 3,928,000,000 randomly generated {\it real} (two-rebit) density matrices, with the sampled Bloch radii $r_A,r_B \in [0,1]$ discretized into intervals of length $\frac{1}{100}$. Our estimate of the lower crossover point is $\tilde{r}_A=0.4722$.}
\end{figure}

\begin{figure}
\includegraphics{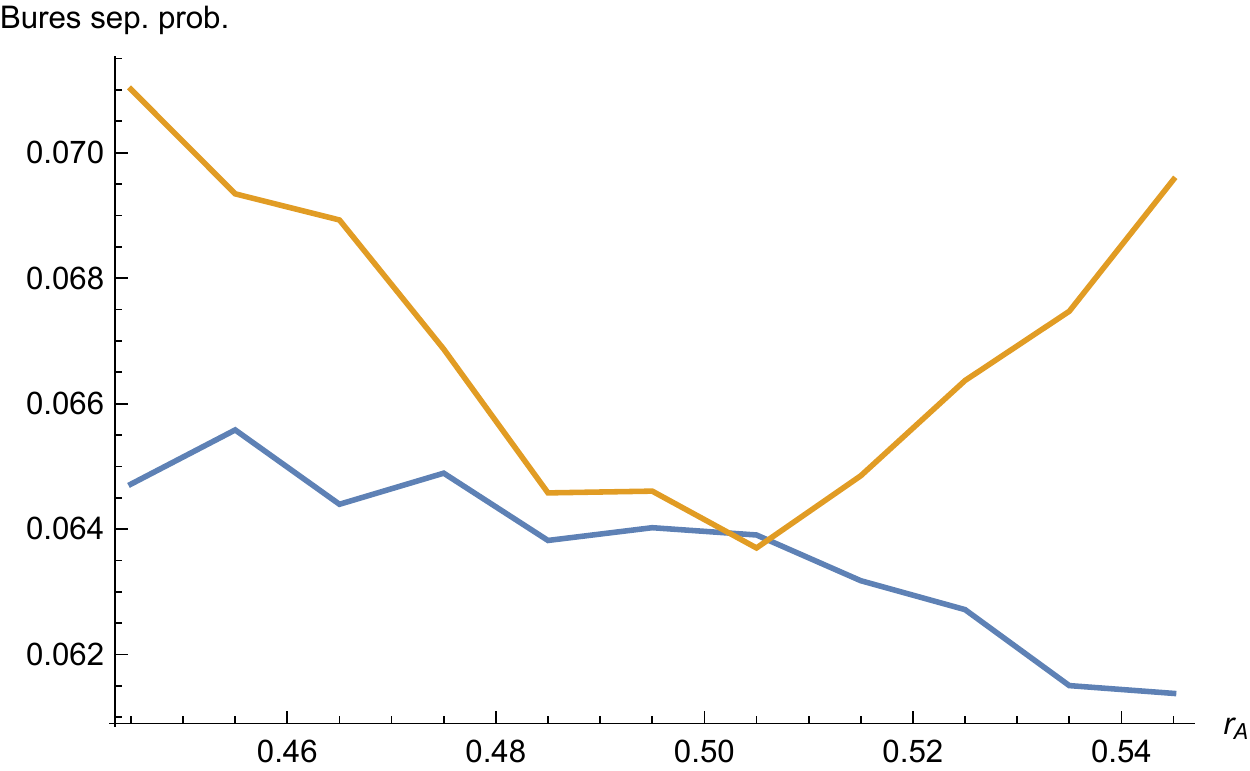}
\caption{\label{fig:RefinedBuresUpperLower}Joint plot of (lower) $p^{Bures}(r_A,r_A)$ and 
(upper) $p^{Bures}(r_A,1-r_A)$ curves based upon 424,000,000 randomly generated density matrices. 
The two curves appear to  cross ever so slightly {\it above} $r_A=\frac{1}{2}$, so there is no evidence of significant crossover behavior in this setting.}
\end{figure}

In \cite{Repulsion}, we had also examined the nature of the separability probabilities
$p(r_A,r_B)$ in the $K=4$ (Hilbert-Schmidt) ``toy'' case with the entries of the density matrix restricted to real values (forming a nine-dimensionalÑ-as opposed to 
fifteen-dimensionalÑ-convex set), and also for the two-qubit states endowed with {\it Bures} (minimal monotone) measure \cite{szBURES,ingemarkarol}. Based upon the samples of random density matrices generated there, we further observe (Fig.~\ref{fig:RefinedRealUpperLower}) crossover behavior (of a ``thin" nature) in the former (two-re[al]bit) case, but, interestingly, none apparently (below $r_A=\frac{1}{2}$) in the Bures instance 
(Fig.~\ref{fig:RefinedBuresUpperLower}). (Let us note that the Bures-based Fig. 31 in \cite{Repulsion} showed highly convincingly that, in strong contrast to the use of Hilbert-Schmidt and random induced measures, the Bures separability probability rapidly decreases 
as $r_A$ increases, rather than remains constant, as for all the other scenarios discussed above.) So, we are inclined to believe that
nontrivial crossover behavior is restricted to the use of Hilbert-Schmidt and  associated random induced measures \cite{Induced}, and that the vague Bures crossover in 
Fig.~\ref{fig:RefinedBuresUpperLower} is purely an insignificant sampling phenomenon.

It very strongly appears in the two-rebit case--in contrast to the integral exponents otherwise so far observed--that both the total and separable volume marginal distributions are now proportional to $(1-r_A^2)^\frac{7}{2}$ (with the consequent constant separability probability over $r_A$, being $\frac{29}{64}$ \cite{slaterJModPhys}).
\section{The case of  two-{\it qutrits}}
\subsection{The role of Casimir invariants}
Our focus here and in \cite{Repulsion} has been on the extension of the two-qubit analyses of Milz and Strunz 
\cite{milzstrunz}--in which they found separability probabilities to be constant over 
the (standard) Bloch radius of qubit subsystems--to a bivariate ($r_A, r_B$) setting. In \cite{Casimir}, we found evidence for another
form of extension. It appears that Hilbert-Schmidt and more generally, random induced separability (and PPT [positive partial transpose]) probabilities are constant, additionally,  over ``generalized Bloch radii" (in group-theoretic terms, square roots of 
{\it quadratic} Casimir invariants) of {\it qutrit} subsystems \cite{Goyal}. Further, constancies appear to continue to hold, as well, over {\it cubic}  Casimir invariants (and, hypothetically, over quartic,..., ones) of reduced higher-dimensional (qudit) states.
\subsection{Hilbert-Schmidt Analysis}
\begin{figure}
\includegraphics{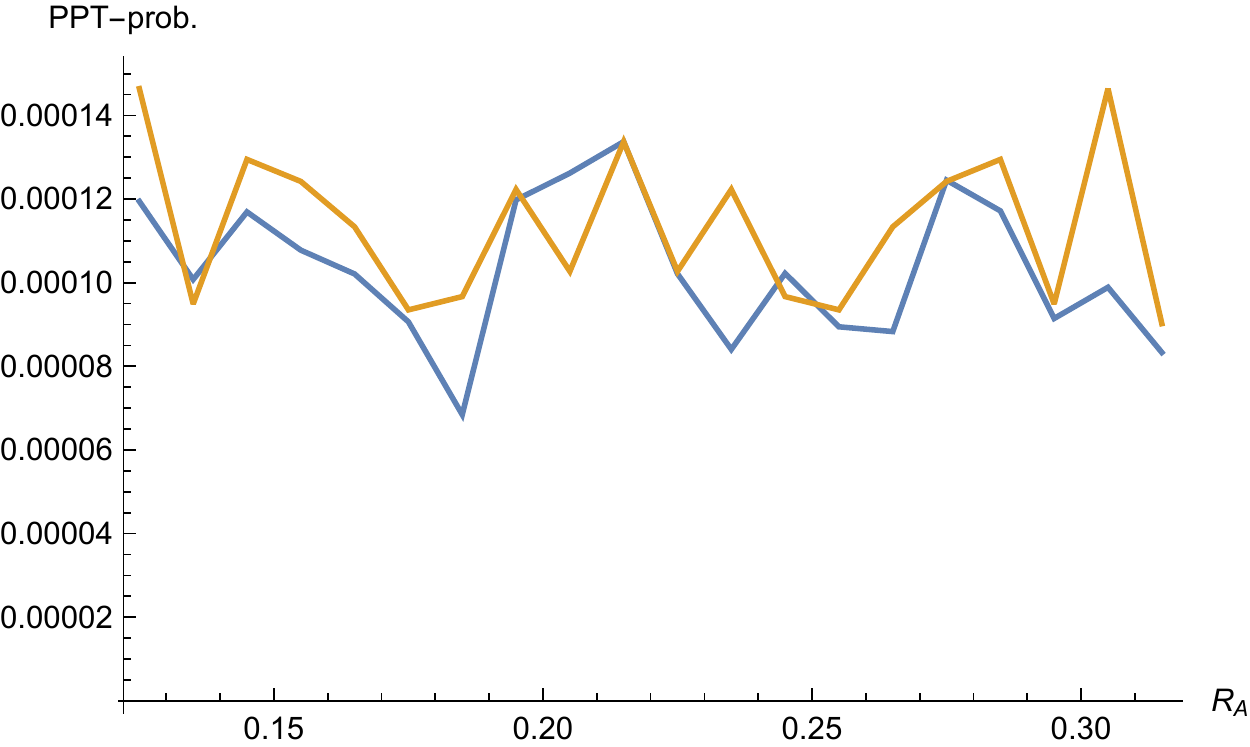}
\caption{\label{fig:TwoQutritHS}The  largely dominant two-{\it qutrit} PPT-probability $p^{Qutrit}_{HS}(R_A,0.435-R_A)$, along with the largely subordinate 
$p^{Qutrit}_{HS}(R_A,R_A)$, based on one hundred million $9 \times 9$ density matrices, randomly generated with respect to Hilbert-Schmidt ($K=9,N=9$) measure. A possible 
crossover region appears near $R_A=0.2$.}
\end{figure}

The question naturally arises of whether or not the various phenomena documented above
in the case of two-qubit systems is also present in some analogous forms in two-{\it qutrit} systems, replacing the standard Bloch radiii ($r_A,r_B$) with their generalized counterparts ($R_A,R_B$).
In \cite[sec. III.A]{Casimir}, one hundred million two-qutrit density matrices were generated,
randomly with respect to Hilbert-Schmidt measure ($K=N=9$). (None of them had 
$R >0.58$.) Only 10,218 of them
had positive partial transposes, with the associated generalized Bloch radii now all lying roughly between 0.05 and 0.44. In Fig.~\ref{fig:TwoQutritHS}, we plot 
the largely dominant $p^{Qutrit}_{HS}(R_A,0.435-R_A)$ curve, along with the 
$p^{Qutrit}_{HS}(R_A,R_A)$ diagonal curve. There is a suggestion of a possible crossover
region near $R_A=0.2$.
\subsection{Random induced ($K=24, N=9$) measure}
\begin{figure}
\includegraphics{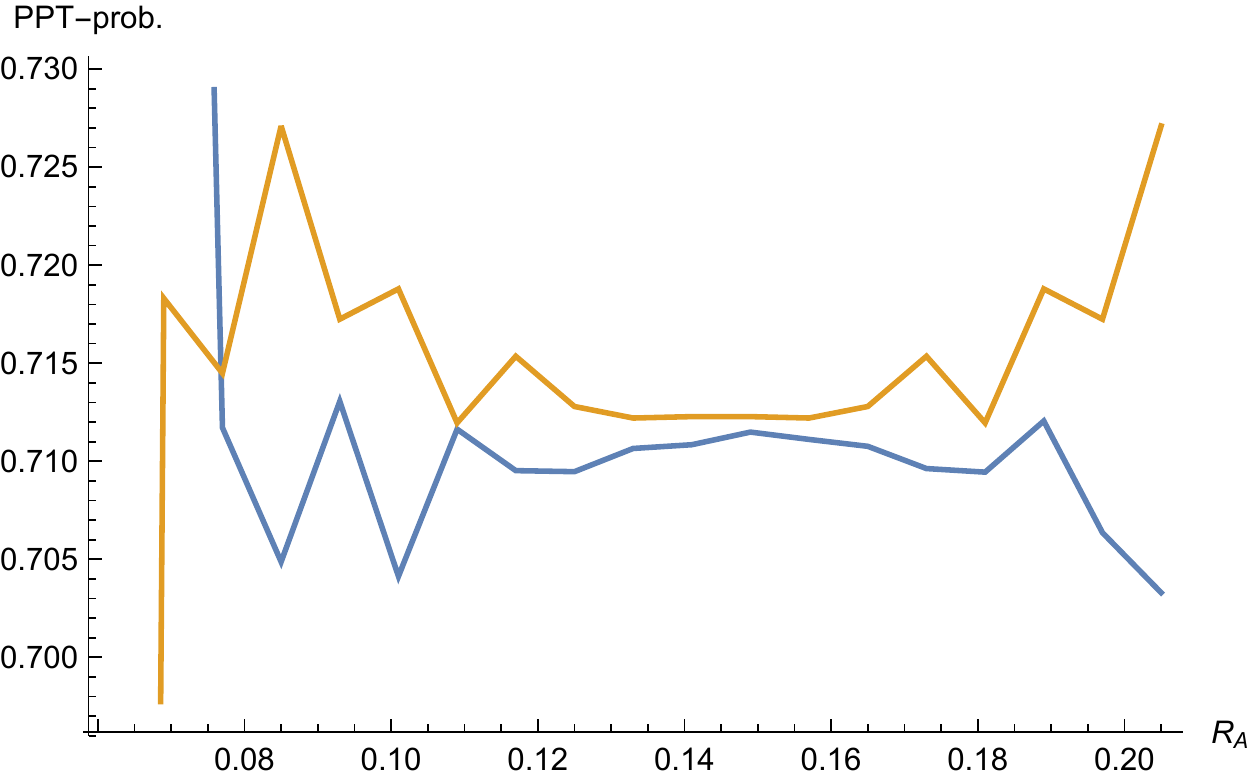}
\caption{\label{fig:TwoQutritK24}The  dominant two-{\it qutrit} PPT-probability $p^{Qutrit}_{K=24}(R_A,0.265-R_A)$, along with the subordinate 
$p^{Qutrit}_{K=24}(R_A,R_A)$, based on 36,400,000 $9 \times 9$ density matrices, randomly generated with respect to induced ($K=24,N=9$) measure. No crossover seems evident.}
\end{figure}

As a supplementary 
exercise---initially being concerned that the previous PPT-probability was too small to detect meaningful effects---we generated 36,400,000 two-qutrit density matrices, with
respect to random induced ($K=24, N=9$) measure. The sample PPT-probability was now,  
orders of magnitude greater than 0.00010218, that is, 0.71179. In  Fig.\ref{fig:TwoQutritK24}, we plot the quasi-antidiagonal $p^{Qutrit}_{K=24}(R_A,0.265-R_A)$ and the diagonal $p^{Qutrit}_{K=24}(R_A,R_A)$ curves.
We see no crossover behavior, noting the restricted range of values of $R_A$, beyond which no significant data were obtained. So, only generalized Bloch radii repulsion--and not attraction--is evident in this plot.
\section{The ``hybrid" qubit-qutrit case} \label{Hybrid}
In \cite[sec. II]{Casimir}, we also conducted a qubit-qutrit analysis based upon
one hundred million $6 \times 6$ density matrices, randomly generated with respect to
Hilbert-Schmidt ($K=N=6$) measure. Let us consider the $A$ subsystem there to be that of the reduced state qubit, and the $B$ subsystem to be that of the reduced state qutrit. Now, we are in a situation where we have no obvious reason to expect that 
the $100 \times 100$ data matrix obtained by using ``bins" of length $\frac{1}{100}$ 
for both $r_A$ and $R_B$ to tend to be symmetric in nature. 

\begin{figure}
\includegraphics{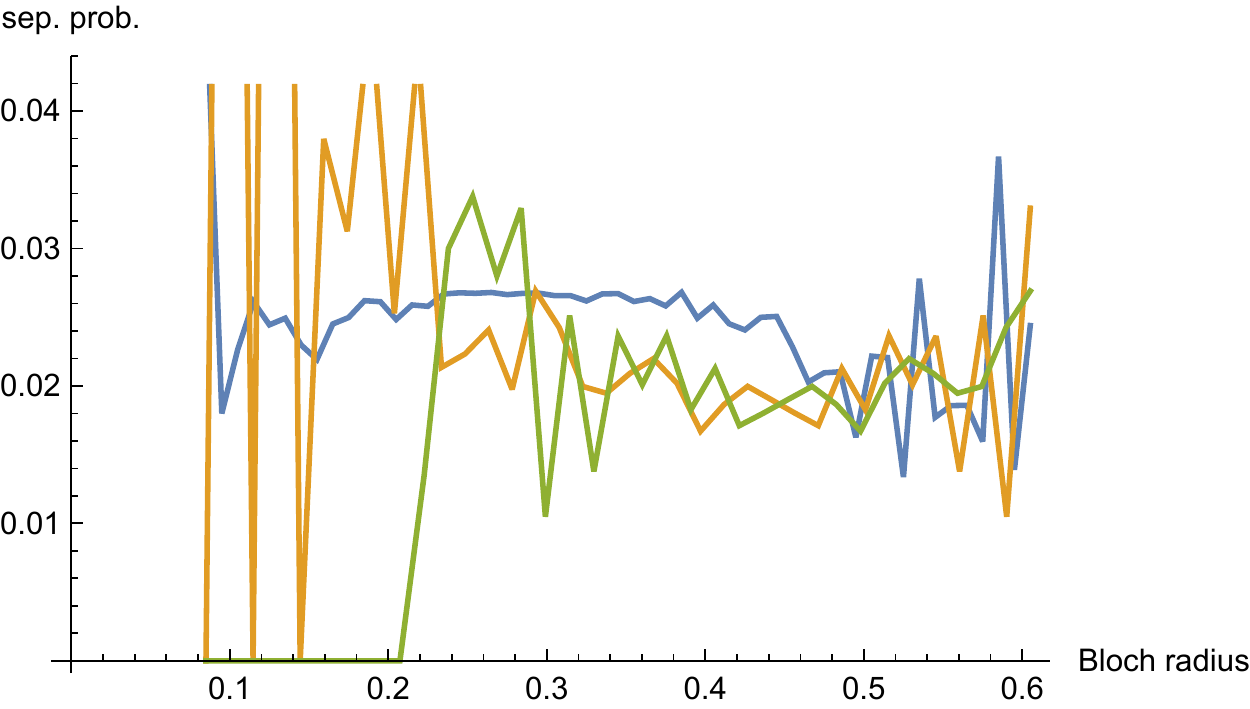}
\caption{\label{fig:QubitQutritTriple}The most level (least jagged) of the three 
qubit-qutrit-based curves
corresponds to the diagonal $p_{HS}^{QubQut}(r_A,r_A)=p_{HS}^{QubQut}(R_B,R_B)$ curve, the most jagged (highest) to the antidiagonal $p_{HS}^{QubQut}(1-R_B,R_B)$ curve and the intermediate one to the antidiagonal reversal $p_{HS}^{QubQut}(r_A,1-r_A)$.}
\end{figure}

In Fig.~\ref{fig:QubitQutritTriple}, we now plot {\it three} curves of interest. The smoothest in character corresponds to the ``diagonal" case, when the qubit Bloch radius ($r_A$) is equal in magnitude (modulo bin size) to the qutrit generalized 
Bloch radius ($R_B$). The most jagged of the three curves is
the antidiagonal $p_{HS}^{QubQut}(1-R_B,R_B)$ one, while  
the intermediate one, $p_{HS}^{QubQut}(r_A,1-r_A)$, is the reversal of that antidiagonal.
The possibility appears of a crossover-type region between 0.3 and 0.5, in which the 
diagonal $r_A=R_B$ curve is dominant.
\subsection{Further possible hybrid analyses}
Additional ``hybrid" analyses such as the qubit-qutrit one just described (sec.~\ref{Hybrid}) were reported in \cite{Casimir}. These included a qubit-qudit ($8 \times 8$ density matrix) analysis \cite[sec. III.B]{Casimir}, as well as two further qubit-qutrit studies. One of these two was based on random induced ($K=9, N=6$), rather than strictly Hilbert-Schmidt, measure \cite[sec. VI]{Casimir}. The other employed the {\it cubic} Casimir invariant (rather than the square root of
the {\it quadratic} invariant--that is the qutrit generalized Bloch radius) \cite[sec. IV.A]{Casimir}.
We might also pursue ``crossover" investigations in these further hybrid settings. Then again, the 
$100 \times 100$ data matrices that were 
generated (by ``binning" the values of the two differing forms of Bloch radii recorded into intervals of length $\frac{1}{100}$) can not be expected to be fundamentally 
symmetric in character.
\section{Two-Qubit ($r_B=\frac{1}{2}$) Analyses}
Aside from the $X$-states $K=4$ (Hilbert-Schmidt) and $K=5$ analyses reported above 
(secs.~\ref{XK=4} and \ref{XK=5}), much work remains to place the other scenarios studied here and similar  ones in a more formal, rigorous  setting. Our results have concentrated on the relations (intersections,\ldots) between ``diagonal" 
and ``antidiagonal" {\it one}-dimensional sections of {\it bivariate} distributions--themselves worthy of fuller understandings. Perhaps the form of one-dimensional section most natural/appealing to study, to yield more insights in addition to these two 
types,  would, in the two-qubit context, be $p(r_A,\frac{1}{2})$, that is setting $r_B=\frac{1}{2}$. We now briefly investigate this issue.

\begin{figure}
\includegraphics{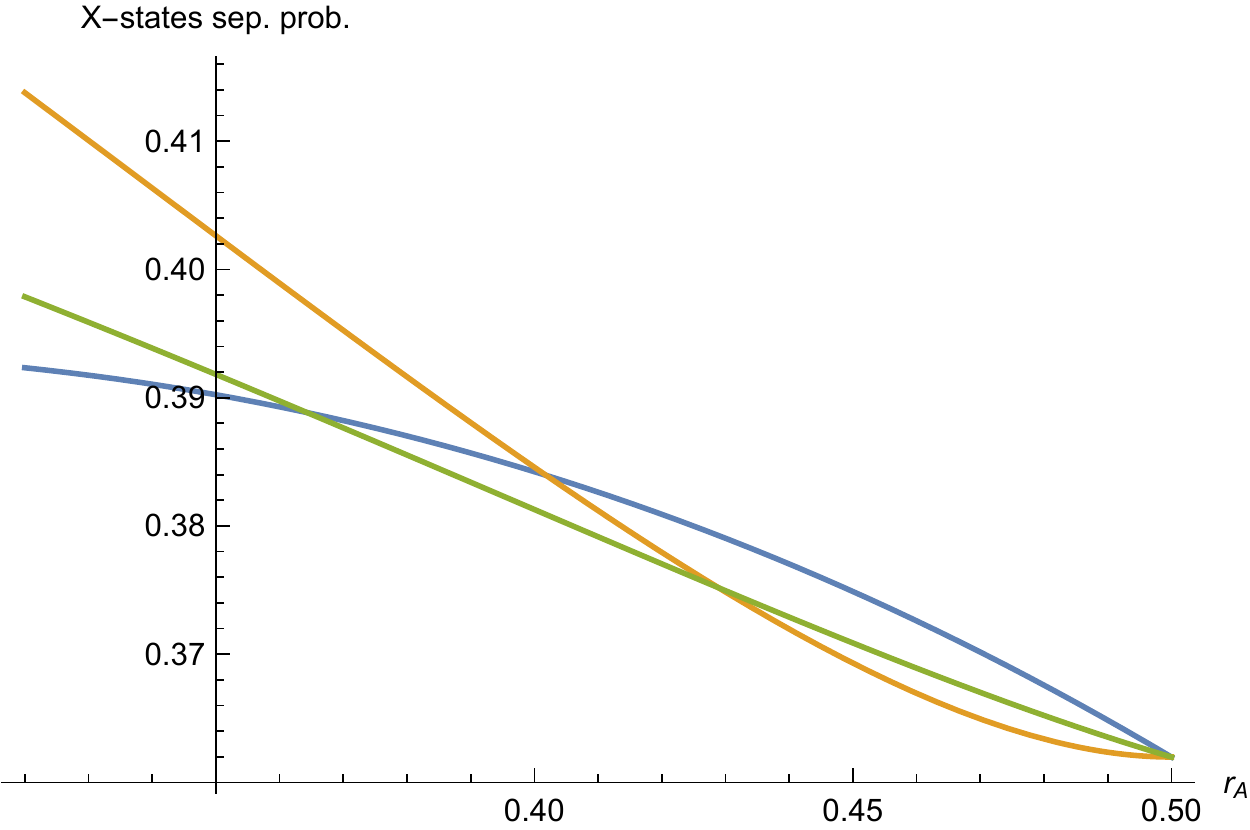}
\caption{\label{fig:ThirdOneDsect}Joint plot for the $X$-states $K=4$ (Hilbert-Schmidt) two-qubit model of the three curves $p^{(X)}_{HS}(r_A,\frac{1}{2})$, $p^{(X)}_{HS}(r_A,r_A)$ and $p^{(X)}_{HS}(r_A,1-r_A)$. The first of these three lies between the other
two near $r_A = \frac{1}{2}$.}
\end{figure}

\begin{figure}
\includegraphics{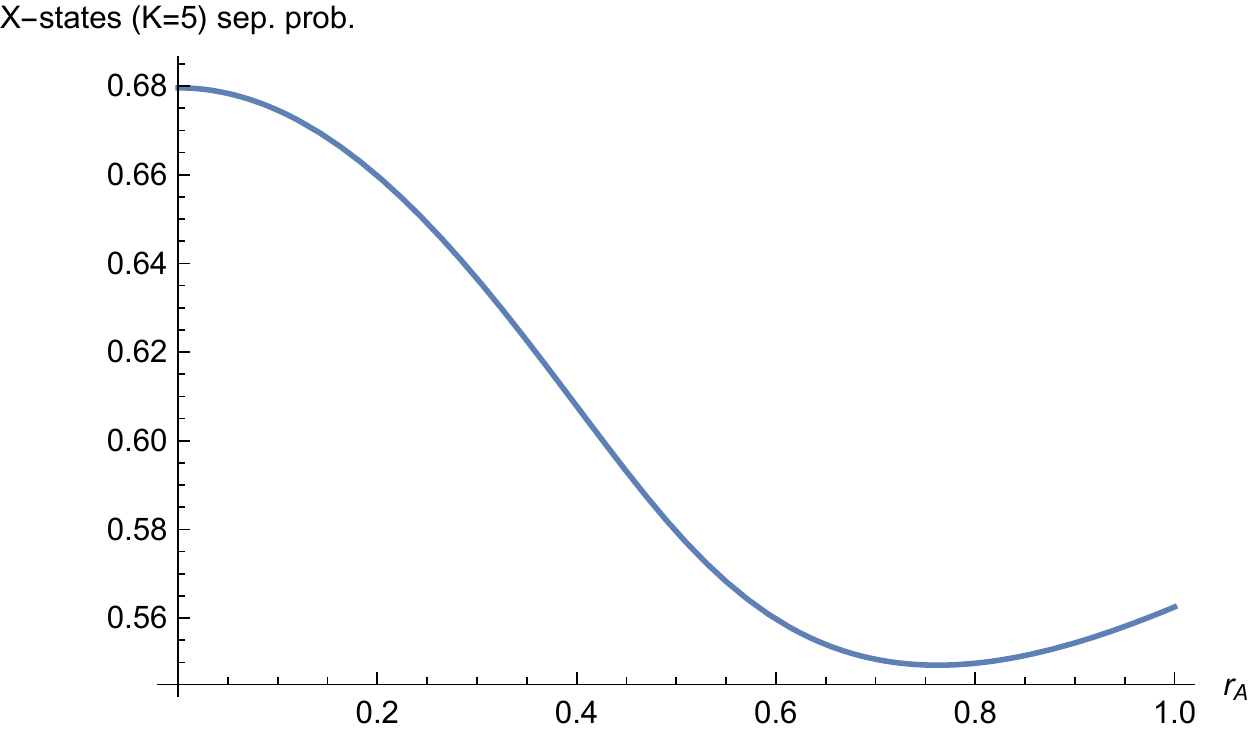}
\caption{\label{fig:ToyK5OneHalf}Plot of 
$p^{(X)}_{K=5}(r_A,\frac{1}{2})$ given by (\ref{XK5onehalf})}
\end{figure}

For our initially studied $X$-states $K=4$ (Hilbert-Schmidt) model (sec.~\ref{XK=4}), we have the result
\begin{equation} \label{onehalfsect}
p^{(X)}_{HS}(r_A,\frac{1}{2})=
\begin{cases}
 \frac{35 r_A^4-50 r_A^2+19}{44-80 r_A^2} & 0<r_A<\frac{1}{2} \\
 \frac{128 r_A^2+29 r_A+23}{32 \left(4 r_A^2+12 r_A-1\right)} &
   \frac{1}{2}<r_A<1 \\
 \frac{139}{384} & r_A=r_B
\end{cases}.
\end{equation}
Expanding upon Figs.~\ref{fig:XbivariateUpperLower} and \ref{fig:RefinedXUpperLower}, in Fig.~\ref{fig:ThirdOneDsect} we plot $p^{(X)}_{HS}(r_A,\frac{1}{2})$, along with the previously jointly plotted $p^{(X)}_{HS}(r_A,r_A)$ and $p^{(X)}_{HS}(r_A,1-r_A)$.
All three curves intersect obviously (by construction) at $r_A =\frac{1}{2}$. Additionally, the first two
listed intersect at $r_A=0.364314$, a root of the quintic equation
\begin{equation}
10 r_A^5+17 r_A^4-24 r_A^3-18 r_A^2+6 r_A+1=0,
\end{equation}
and the first and the third at $r_A=0.428908$, a root of the sextic equation
\begin{equation}
10 r_A^6-7 r_A^5-34 r_A^4-6 r_A^3+30 r_A^2+5 r_A-6=0.
\end{equation}
For the $X$-states $K=5$ model (sec.~\ref{XK=5}) (Fig.~\ref{fig:ToyK5OneHalf}), we have
\begin{equation} \label{XK5onehalf}
p^{(X)}_{K=5}(r_A,\frac{1}{2})=
\begin{cases}
 \frac{-231 r_A^6+441 r_A^4-297 r_A^2+70}{336 r_A^4-360
   r_A^2+103} & 0<r_A<\frac{1}{2} \\
 \frac{512 r_A^4+2560 r_A^3-384 r_A^2+679 r_A+35}{32 \left(16
   r_A^4+80 r_A^3+120 r_A^2-40 r_A+13\right)} &
   \frac{1}{2}<r_A<1 \\
 \frac{1261}{2176} & r_A=r_B
\end{cases}.
\end{equation}
\section{Concluding Remarks}
If we examine the analytically-derived total and separable volume {\it piecewise} formulas 
((\ref{Xtotal}), (\ref{Xsep})) for the $X$-states ($K=4$, Hilbert-Schmidt) ``toy" model 
that we have employed as our starting point, we see that the pieces are 
bivariate {\it polynomials} in $r_A$ and $r_B$. On the other hand, the  
analogous pieces in the candidate (well-fitting) formulas ((\ref{totalK3}),  (\ref{separableK3})) we have advanced in the 15-dimensional $K=3$ case, are such polynomials {\it divided} by $r_A$ or $r_B$--that is, {\it rational} functions. A similar situation holds with regard to our working formulas for the 15-dimensional $K=4$ (Hilbert-Schmidt) scenario--which we have employed for our estimate (\ref{K4antidiagonal})
of $p^{K=4}_{HS}(r_A,1-r_A)$. This type of functional difference is a matter of some interest/concern, meriting further investigation. (The distinction between rational and polynomial functions, of course,  disappears in the computation of the {\it ratios} yielding the separability probabilities.)
These volume formulas were constructed so as to satisfy the marginal constraints $(1-r_A^2)^n$, 
$n=2 (K-1)$, $K=3,4$, with the resultant indicated proportionalities of $\frac{1}{14}$ and $\frac{8}{33}$,
and also to satisfy the apparent diagonal separability probabilities (\ref{K3diagonal})  and (\ref{K4diagonal}) of $p_{K=3}(r_A,r_A)= \frac{1}{32} \left(1-r_A\right) \left(r_A^2+6 r_A+1\right)$ and $p^{K=4}_{HS}(r_A,r_A) = -\frac{35 \left(r_A-1\right) \left(58 r_A^2+17 r_A+2\right)}{384 \left(8 r_A+1\right)}$, respectively.

To each bin of length $\frac{1}{100}$ employed to discretize our computations, we have 
simply attributed a Bloch radius equal to the midpoint of the bin. Perhaps, one can utilize the data themselves to assign values to the bins that would lead to more accurate volume and probability estimations.

It, of course, would be desirable to analytically derive total and separable volume
and (consequent) separability probability formulas for the full range of scenarios considered above.
To this point in time, we are aware of only one broadly successful formal endeavor in this general direction.
By this, we mean the work of  Szarek, Bengtsson and {\.Z}yczkowski, in which they were able to establish 
that the Hilbert-Schmidt separability (and, more generally, \newline PPT-) probabilities of boundary states, corresponding to minimally degenerate density matrices (those with exactly one zero eigenvalue),  are one-half of the corresponding probabilities of generic nondegenerate  density matrices \cite{sbz}.

In a most interesting recent development, Mark Fischler has given a highly detailed response to a question I posed on the physics stack exchange, as to the possibility of constructing ``bivariate symmetric (polynomial) Hilbert-Schmidt two-qubit volume functions over the unit square with certain properties''. The interchange can be found at 
http://physics.stackexchange.com/questions/201369/construct-bivariate-symmetric-polynomial-hilbert-schmidt-two-qubit-volume-func.

\bibliography{Domains5}

\begin{thebibliography}{29}
\expandafter\ifx\csname natexlab\endcsname\relax\def\natexlab#1{#1}\fi
\expandafter\ifx\csname bibnamefont\endcsname\relax
  \def\bibnamefont#1{#1}\fi
\expandafter\ifx\csname bibfnamefont\endcsname\relax
  \def\bibfnamefont#1{#1}\fi
\expandafter\ifx\csname citenamefont\endcsname\relax
  \def\citenamefont#1{#1}\fi
\expandafter\ifx\csname url\endcsname\relax
  \def\url#1{\texttt{#1}}\fi
\expandafter\ifx\csname urlprefix\endcsname\relax\def\urlprefix{URL }\fi
\providecommand{\bibinfo}[2]{#2}
\providecommand{\eprint}[2][]{\url{#2}}

\bibitem[{\citenamefont{Jevtic et~al.}(2014)\citenamefont{Jevtic, Pusey,
  Jennings, and Rudolph}}]{Jevtic}
\bibinfo{author}{\bibfnamefont{S.}~\bibnamefont{Jevtic}},
  \bibinfo{author}{\bibfnamefont{M.}~\bibnamefont{Pusey}},
  \bibinfo{author}{\bibfnamefont{D.}~\bibnamefont{Jennings}}, \bibnamefont{and}
  \bibinfo{author}{\bibfnamefont{T.}~\bibnamefont{Rudolph}},
  \bibinfo{journal}{Phys. Rev. Lett.} \textbf{\bibinfo{volume}{113}},
  \bibinfo{pages}{020402} (\bibinfo{year}{2014}).

\bibitem[{\citenamefont{Milz and Strunz}(2015)}]{milzstrunz}
\bibinfo{author}{\bibfnamefont{S.}~\bibnamefont{Milz}} \bibnamefont{and}
  \bibinfo{author}{\bibfnamefont{W.~T.} \bibnamefont{Strunz}},
  \bibinfo{journal}{J. Phys. A} \textbf{\bibinfo{volume}{48}},
  \bibinfo{pages}{035306} (\bibinfo{year}{2015}).

\bibitem[{\citenamefont{Slater}(2015)}]{Repulsion}
\bibinfo{author}{\bibfnamefont{P.~B.} \bibnamefont{Slater}},
  \bibinfo{journal}{arXiv preprint arXiv:1506.08739}  (\bibinfo{year}{2015}).

\bibitem[{\citenamefont{Gamel}(2016)}]{Gamel}
\bibinfo{author}{\bibfnamefont{O.}~\bibnamefont{Gamel}},
  \bibinfo{journal}{Phys. Rev. A} \textbf{\bibinfo{volume}{93}},
  \bibinfo{pages}{062320} (\bibinfo{year}{2016}).

\bibitem[{\citenamefont{Mendon\c{c}a et~al.}(2014)\citenamefont{Mendon\c{c}a,
  Marchiolli, and Galetti}}]{Xstates2}
\bibinfo{author}{\bibfnamefont{P.}~\bibnamefont{Mendon\c{c}a}},
  \bibinfo{author}{\bibfnamefont{M.~A.} \bibnamefont{Marchiolli}},
  \bibnamefont{and} \bibinfo{author}{\bibfnamefont{D.}~\bibnamefont{Galetti}},
  \bibinfo{journal}{Ann. Phys.} \textbf{\bibinfo{volume}{351}},
  \bibinfo{pages}{79} (\bibinfo{year}{2014}).

\bibitem[{\citenamefont{Slater}(2007)}]{slater833}
\bibinfo{author}{\bibfnamefont{P.~B.} \bibnamefont{Slater}},
  \bibinfo{journal}{J. Phys. A} \textbf{\bibinfo{volume}{40}},
  \bibinfo{pages}{14279} (\bibinfo{year}{2007}).

\bibitem[{\citenamefont{Slater and Dunkl}(2012)}]{MomentBased}
\bibinfo{author}{\bibfnamefont{P.~B.} \bibnamefont{Slater}} \bibnamefont{and}
  \bibinfo{author}{\bibfnamefont{C.~F.} \bibnamefont{Dunkl}},
  \bibinfo{journal}{J. Phys. A} \textbf{\bibinfo{volume}{45}},
  \bibinfo{pages}{095305} (\bibinfo{year}{2012}).

\bibitem[{\citenamefont{Slater}(2013)}]{slaterJModPhys}
\bibinfo{author}{\bibfnamefont{P.~B.} \bibnamefont{Slater}},
  \bibinfo{journal}{J. Phys. A} \textbf{\bibinfo{volume}{46}},
  \bibinfo{pages}{445302} (\bibinfo{year}{2013}).

\bibitem[{\citenamefont{Fei and Joynt}()}]{FeiJoynt}
\bibinfo{author}{\bibfnamefont{J.}~\bibnamefont{Fei}} \bibnamefont{and}
  \bibinfo{author}{\bibfnamefont{R.}~\bibnamefont{Joynt}},
  \eprint{arXiv.1409:1993}.

\bibitem[{\citenamefont{Slater and Dunkl}(2015{\natexlab{a}})}]{WholeHalf}
\bibinfo{author}{\bibfnamefont{P.~B.} \bibnamefont{Slater}} \bibnamefont{and}
  \bibinfo{author}{\bibfnamefont{C.~F.} \bibnamefont{Dunkl}},
  \bibinfo{journal}{J. Geom. Phys.} \textbf{\bibinfo{volume}{90}},
  \bibinfo{pages}{42} (\bibinfo{year}{2015}{\natexlab{a}}).

\bibitem[{\citenamefont{Khvedelidze and Rogojin}(2015)}]{Dubna}
\bibinfo{author}{\bibfnamefont{A.}~\bibnamefont{Khvedelidze}} \bibnamefont{and}
  \bibinfo{author}{\bibfnamefont{I.}~\bibnamefont{Rogojin}},
  \bibinfo{journal}{Zap. Nauchn. Sem. POMI} \textbf{\bibinfo{volume}{432}},
  \bibinfo{pages}{274} (\bibinfo{year}{2015}).

\bibitem[{\citenamefont{Fonseca-Romero
  et~al.}(2012)\citenamefont{Fonseca-Romero, Martinez-Rinc{\'o}n, and
  Viviescas}}]{Fonseca-Romero}
\bibinfo{author}{\bibfnamefont{K.~M.} \bibnamefont{Fonseca-Romero}},
  \bibinfo{author}{\bibfnamefont{J.~M.} \bibnamefont{Martinez-Rinc{\'o}n}},
  \bibnamefont{and}
  \bibinfo{author}{\bibfnamefont{C.}~\bibnamefont{Viviescas}},
  \bibinfo{journal}{Phys. Rev. A} \textbf{\bibinfo{volume}{86}},
  \bibinfo{pages}{042325} (\bibinfo{year}{2012}).

\bibitem[{\citenamefont{Shang et~al.}(2015)\citenamefont{Shang, Seah, K.Ng,
  Nott, and Englert}}]{Shang}
\bibinfo{author}{\bibfnamefont{J.}~\bibnamefont{Shang}},
  \bibinfo{author}{\bibfnamefont{Y.-L.} \bibnamefont{Seah}},
  \bibinfo{author}{\bibfnamefont{H.}~\bibnamefont{K.Ng}},
  \bibinfo{author}{\bibfnamefont{D.~J.} \bibnamefont{Nott}}, \bibnamefont{and}
  \bibinfo{author}{\bibfnamefont{B.-T.} \bibnamefont{Englert}},
  \bibinfo{journal}{New J. Phys.} \textbf{\bibinfo{volume}{17}},
  \bibinfo{pages}{043017} (\bibinfo{year}{2015}).

\bibitem[{\citenamefont{Lovas and Andai}(2016)}]{lovasandai}
\bibinfo{author}{\bibfnamefont{A.}~\bibnamefont{Lovas}} \bibnamefont{and}
  \bibinfo{author}{\bibfnamefont{A.}~\bibnamefont{Andai}},
  \bibinfo{journal}{arXiv preprint arXiv:1610.01410}  (\bibinfo{year}{2016}).

\bibitem[{\citenamefont{Slater and
  Dunkl}(2015{\natexlab{b}})}]{LatestCollaboration}
\bibinfo{author}{\bibfnamefont{P.~B.} \bibnamefont{Slater}} \bibnamefont{and}
  \bibinfo{author}{\bibfnamefont{C.~F.} \bibnamefont{Dunkl}},
  \bibinfo{journal}{Adv. Math. Phys.} \textbf{\bibinfo{volume}{2015}},
  \bibinfo{pages}{621353} (\bibinfo{year}{2015}{\natexlab{b}}).

\bibitem[{\citenamefont{{\.Z}yczkowski and Sommers}(2001)}]{Induced}
\bibinfo{author}{\bibfnamefont{K.}~\bibnamefont{{\.Z}yczkowski}}
  \bibnamefont{and} \bibinfo{author}{\bibfnamefont{H.-J.}
  \bibnamefont{Sommers}}, \bibinfo{journal}{J. Phys. A}
  \textbf{\bibinfo{volume}{A34}}, \bibinfo{pages}{7111} (\bibinfo{year}{2001}).

\bibitem[{\citenamefont{Aubrun et~al.}(2014)\citenamefont{Aubrun, Szarek, and
  Ye}}]{aubrun2}
\bibinfo{author}{\bibfnamefont{G.}~\bibnamefont{Aubrun}},
  \bibinfo{author}{\bibfnamefont{S.~J.} \bibnamefont{Szarek}},
  \bibnamefont{and} \bibinfo{author}{\bibfnamefont{D.}~\bibnamefont{Ye}},
  \bibinfo{journal}{Commun. Pure Appl. Math.} \textbf{\bibinfo{volume}{LXVII}},
  \bibinfo{pages}{0129} (\bibinfo{year}{2014}).

\bibitem[{\citenamefont{Adachi et~al.}(2009)\citenamefont{Adachi, Toda, and
  Kubotani}}]{adachi2009random}
\bibinfo{author}{\bibfnamefont{S.}~\bibnamefont{Adachi}},
  \bibinfo{author}{\bibfnamefont{M.}~\bibnamefont{Toda}}, \bibnamefont{and}
  \bibinfo{author}{\bibfnamefont{H.}~\bibnamefont{Kubotani}},
  \bibinfo{journal}{Annals of Physics} \textbf{\bibinfo{volume}{324}},
  \bibinfo{pages}{2278} (\bibinfo{year}{2009}).

\bibitem[{\citenamefont{Braga et~al.}(2010)\citenamefont{Braga, Souza, and
  Mizrahi}}]{BSM}
\bibinfo{author}{\bibfnamefont{H.}~\bibnamefont{Braga}},
  \bibinfo{author}{\bibfnamefont{S.}~\bibnamefont{Souza}}, \bibnamefont{and}
  \bibinfo{author}{\bibfnamefont{S.~S.} \bibnamefont{Mizrahi}},
  \bibinfo{journal}{Phys. Rev. A} \textbf{\bibinfo{volume}{81}},
  \bibinfo{pages}{042310} (\bibinfo{year}{2010}).

\bibitem[{\citenamefont{Sungur and Ng}(2005)}]{SungurNg}
\bibinfo{author}{\bibfnamefont{E.~A.} \bibnamefont{Sungur}} \bibnamefont{and}
  \bibinfo{author}{\bibfnamefont{P.}~\bibnamefont{Ng}},
  \bibinfo{journal}{Commun. Stat.--Theory and Methods}
  \textbf{\bibinfo{volume}{34}}, \bibinfo{pages}{2269} (\bibinfo{year}{2005}).

\bibitem[{\citenamefont{Ruschendorf et~al.}(1996)\citenamefont{Ruschendorf,
  Schweizer, and Taylor}}]{Ruschendorf}
\bibinfo{author}{\bibfnamefont{L.}~\bibnamefont{Ruschendorf}},
  \bibinfo{author}{\bibfnamefont{B.}~\bibnamefont{Schweizer}},
  \bibnamefont{and} \bibinfo{author}{\bibfnamefont{M.~D.}
  \bibnamefont{Taylor}}, \emph{\bibinfo{title}{Distributions with fixed
  marginals and related topics}} (\bibinfo{publisher}{Institute of Mathematical
  Statistics}, \bibinfo{address}{Hayward, CA}, \bibinfo{year}{1996}).

\bibitem[{\citenamefont{de~Vicente}()}]{deVicente}
\bibinfo{author}{\bibfnamefont{J.~I.} \bibnamefont{de~Vicente}},
  \emph{\bibinfo{title}{Further results on entanglement detection and
  quantificiation from the correlation matrix criterion}},
  \eprint{arXiv:0705.2583}.

\bibitem[{\citenamefont{Bengtsson and {\.Z}yczkowski}(2006)}]{ingemarkarol}
\bibinfo{author}{\bibfnamefont{I.}~\bibnamefont{Bengtsson}} \bibnamefont{and}
  \bibinfo{author}{\bibfnamefont{K.}~\bibnamefont{{\.Z}yczkowski}},
  \emph{\bibinfo{title}{Geometry of Quantum States}}
  (\bibinfo{publisher}{Cambridge}, \bibinfo{address}{Cambridge},
  \bibinfo{year}{2006}).

\bibitem[{\citenamefont{Miszczak}(2012)}]{Miszczak}
\bibinfo{author}{\bibfnamefont{J.~A.} \bibnamefont{Miszczak}},
  \bibinfo{journal}{Comput. Phys. Commun.} \textbf{\bibinfo{volume}{183}},
  \bibinfo{pages}{118} (\bibinfo{year}{2012}).

\bibitem[{\citenamefont{Miszczak}(2013)}]{Miszczak2}
\bibinfo{author}{\bibfnamefont{J.~A.} \bibnamefont{Miszczak}},
  \bibinfo{journal}{Comput. Phys. Commun.} \textbf{\bibinfo{volume}{184}},
  \bibinfo{pages}{257} (\bibinfo{year}{2013}).

\bibitem[{\citenamefont{Sommers and {\.Z}yczkowski}(2003)}]{szBURES}
\bibinfo{author}{\bibfnamefont{H.-J.} \bibnamefont{Sommers}} \bibnamefont{and}
  \bibinfo{author}{\bibfnamefont{K.}~\bibnamefont{{\.Z}yczkowski}},
  \bibinfo{journal}{J. Phys. A} \textbf{\bibinfo{volume}{36}},
  \bibinfo{pages}{10083} (\bibinfo{year}{2003}).

\bibitem[{\citenamefont{Slater}(2016)}]{Casimir}
\bibinfo{author}{\bibfnamefont{P.~B.} \bibnamefont{Slater}},
  \bibinfo{journal}{Quantum Information Processing} pp. \bibinfo{pages}{1--16}
  (\bibinfo{year}{2016}).

\bibitem[{\citenamefont{Goyal et~al.}(2016)\citenamefont{Goyal, Simon, Singh,
  and Simon}}]{Goyal}
\bibinfo{author}{\bibfnamefont{S.~K.} \bibnamefont{Goyal}},
  \bibinfo{author}{\bibfnamefont{B.~N.} \bibnamefont{Simon}},
  \bibinfo{author}{\bibfnamefont{R.}~\bibnamefont{Singh}}, \bibnamefont{and}
  \bibinfo{author}{\bibfnamefont{S.}~\bibnamefont{Simon}}, \bibinfo{journal}{J.
  Phys. A} \textbf{\bibinfo{volume}{49}}, \bibinfo{pages}{165203}
  (\bibinfo{year}{2016}).

\bibitem[{\citenamefont{Szarek et~al.}(2006)\citenamefont{Szarek, Bengtsson,
  and {\.Z}yczkowski}}]{sbz}
\bibinfo{author}{\bibfnamefont{S.}~\bibnamefont{Szarek}},
  \bibinfo{author}{\bibfnamefont{I.}~\bibnamefont{Bengtsson}},
  \bibnamefont{and}
  \bibinfo{author}{\bibfnamefont{K.}~\bibnamefont{{\.Z}yczkowski}},
  \bibinfo{journal}{J. Phys. A} \textbf{\bibinfo{volume}{39}},
  \bibinfo{pages}{L119} (\bibinfo{year}{2006}).

\end{thebibliography}

\end{document}